# Trading Privacy for the Greater Social Good: How Did America React During COVID-19?

Anindya Ghose[a]
Beibei Li[b]
Meghanath Macha[b]
Chenshuo Sun[a]
Natasha Ying Zhang Foutz[c]

[a] New York University, Stern School of Business
[b] Carnegie Mellon University, Heinz College, Information and Public Policy Management
[c] University of Virginia, McIntire School of Commerce

Working Paper, *June 9, 2020*[1]
For the most recent version, please check [here](here).

**Abstract**

Digital contact tracing and analysis of social distancing from smartphone location data are two prime examples of non-therapeutic interventions used in many countries to mitigate the impact of the COVID-19 pandemic. While many understand the importance of trading personal privacy for the public good, others have been alarmed at the potential for surveillance via measures enabled through location tracking on smartphones. In our research, we analyzed massive yet atomic individual-level location data containing over 22 billion records from ten "Blue" (Democratic) and ten "Red" (Republican) cities in the U.S., based on which we present, herein, some of the first evidence of how Americans responded to the increasing concerns that government authorities, the private sector, and public health experts might use individual-level location data to track the COVID-19 spread. First, we found a significant decreasing trend of mobile-app location-sharing opt out. Whereas areas with more Democrats were more privacy-concerned than areas with more Republicans before the advent of the COVID-19 pandemic, there was a significant decrease in the overall opt-out rates after COVID-19, and this effect was more salient among Democratic than Republican cities. Second, people who practiced social distancing (i.e., those who traveled less and interacted with fewer close contacts during the pandemic) were also less likely to opt out, whereas the converse was true for people who practiced less social-distancing. This relationship also was more salient among Democratic than Republican cities. Third, high-income populations and males, compared with low-income populations and females, were more privacy-conscientious and more likely to opt out of location tracking. Overall, our findings demonstrate that during COVID-19, people in both Blue and Red cities generally reacted in a consistent manner in trading their personal privacy for the greater social good but diverged in the extent of that trade-off along the lines of political affiliation, social-distancing compliance, and demographics.

[1] We thank Panos Adamopolous, Ravi Bapna, Indranil Bardhan, Gordon Burtch, Prithwiraj Choudhury, Alok Gupta, De Liu, Eitan Muller, Unnati Narang, Arun Rai, Danny Sokol, Vilma Todri, Catherine Tucker, Raluca Ursu, Yuqian Xu, Michael F Zhao, Peter Zubscek and webinar participants in the Carlson Online MIS Seminar for very helpful comments that have improved the paper.

# 1. Introduction

In the very early days of COVID-19, as health experts realized that human-to-human transmission was happening, they advised the public to practice "social distancing," a tactic commonly employed in the past to combat health epidemics. Stay-at-home orders were also put in place in many communities globally as another non-pharmaceutical intervention or measure taken to mitigate the COVID-19 spread. The related issue of location tracking of citizens' whereabouts, widely covered by major news outlets, has seized public attention and reignited debate over the efficacy and legality of government surveillance and privacy rights.

In early March of 2020, as COVID-19 continued to spread rapidly all over the world, scientists and health officials were puzzled by the exceedingly high infection and fatality rates in certain parts of the world such as Italy and Spain. Vodafone subsequently provided Italian officials with anonymized customer data to track and analyze population movements in Italy, where a government-mandated lockdown was in place. A crucial insight gleaned from the analysis of the data was that up to 40% of residents in Milan still moved every day 300-500 meters beyond their homes despite the lockdown, which finding revealed the extent of social-distancing non-compliance as a significant driving factor of COVID-19's diffusion in Italy. In the meantime, active contact tracing as well as monitoring of possibly infected people and those who had come into contact with any suspected cases was widely publicized as being effective in curbing the COVID-19 spread,[2] particularly in countries such as Taiwan, China, Singapore, Israel, and South Korea.

Highly granular location data revealing all of the locations of a given consumer in the immediate past is needed for successful implementation of contact tracing. By virtue of the fact that consumers are wedded to their smartphones and have widely adopted wearable technologies, such data is available to telecom providers, digital platforms, wearable technology and smartwatch firms, and mobile app developers.[3] The consequent news headlines in mainstream media such as CNN[4] and Fox News[5] led to many people becoming aware of contact tracing, location-data tracking, and potential surveillance concerns. One of the inadvertent consequences of this heightened awareness was increased public concern about surveillance and privacy.[6] It is well known that apps make it feasible for different elements of the mobile ecosystem to track consumers' locations. One perspective held by some consumers is that since lives are at stake, it is imperative that we look at data privacy through a different lens, at least until the pandemic is mitigated and suppressed. In the UK, more than two-thirds of the population backed the use of CCTV footage, mobile phone data, and credit card records in a mass 'contact tracing' exercise to prevent a second wave of coronavirus infections.[7] On the other hand, skeptical consumers have complained that access to such atomic consumer data is an infringement of their civil liberties. They are especially concerned that contact tracing may enable intrusions into their everyday lives by governments or big tech companies.[8]

---

[2] https://www.yalejreg.com/nc/unlocking-platform-technology-to-combat-health-pandemics-by-anindya-ghose-and-d-daniel-sokol/.
[3] https://mitpress.mit.edu/books/tap.
[4] https://www.cnn.com/2020/03/10/health/coronavirus-contact-tracing/index.html; https://www.cnn.com/2020/03/18/tech/us-government-location-data-coronavirus/index.html.
[5] https://www.foxnews.com/tech/us-government-big-tech-smartphone-coronavirus-google-facebook. https://www.foxnews.com/tech/taiwans-so-called-electronic-fence-monitor-for-those-quarantined-raises-privacy-concerns-report
[6] https://www.ft.com/content/005ab1a8-1691-4e7b-8e10-0d3d2614a276.

A 2017 Reuters poll[9] found that a majority of Americans were unwilling to give up privacy to assist the U.S. government's national security initiatives: over 76% reported being unwilling to yield information on their internet activities even if it would help the U.S. government's security initiatives. But would Americans behave differently during a pandemic? According to April 2020 research from CodeFuel[10], 84% of U.S. adults said that they would be willing to share their health data to deter the spread of the virus, and more than half (58%) of respondents said that they would be open to disclosing their location—both important factors that can help public officials flag hot spots. Another recent survey[11] based on responses from 1,374 American adults found that two-thirds were willing to install an app that would help slow the spread of the virus and reduce the lockdown period, even if that app collected information about their location data and health status. Moreover, people whom the Centers for Disease Control (CDC) has identified as being at higher risk, those who are younger and those who are more technologically savvy, were more likely to be willing to install such an app.

Meanwhile, studies have also shown that whereas low-income populations in the U.S. are aware of a range of digital privacy harms, it is difficult for them to access the tools and strategies that could help them protect their personal information online.[12] Conversely, high-income populations are more likely to be both aware of potential harms and technologically savvier in their ability to protect their data privacy. Besides, studies have shown that political party affiliation and political ideology impact how Americans feel about institutional surveillance more than do income, age, gender, and race.[13] More specifically, a 2018 study showed that Republicans tend to feel pleased about tracking, both online and in real life, while Democrats often feel bad about it.[14] Conservative Republicans were found to have warmer responses to surveillance scenarios, whereas liberal Democrats had cooler responses, with moderates and Independents somewhere in the middle. These facts motivated us to incorporate differences between Democrats and Republicans in their privacy choices as exhibited by differences in opt-out rates on mobile apps that allow for location tracking.

---

[7] https://www.telegraph.co.uk/politics/2020/04/04/public-backs-tracking-people-phones-monitor-coronavirus-infection/.
[8] https://www.latimes.com/politics/story/2020-04-26/privacy-americans-trade-off-trace-coronavirus-contacts.
[9] Dustin Volz, "Most Americans unwilling to give up privacy to thwart attacks: Reuters/Ipsos Poll," Reuters, April 4, 2017, https://www.reuters.com/article/us-usa-cyber-poll/most-americansunwilling-to-give-up-privacy-to-thwart-attacks-reuters-ipsos-poll-idUSKBN1762TQ .
[10] https://www.emarketer.com/content/consumers-are-more-willing-to-share-private-data-during-covid-19
[11] https://blogs.scientificamerican.com/observations/will-americans-be-willing-to-install-covid-19-tracking-apps/
[12] https://datasociety.net/pubs/prv/DataAndSociety_PrivacySecurityandDigitalInequalityPressRelease.pdf.
[13] "We Feel: Partisan Politics Drive Americans' Emotions Regarding Surveillance of Low-Income Populations," https://www.asc.upenn.edu/sites/default/files/documents/Turow-Divided-Final.pdf.
[14] https://www.nytimes.com/2018/04/30/technology/privacy-concerns-politics.html.

In this research, we aimed to examine and quantify the impact of socio-economic factors including demographics (such as age, income, race, gender, and income) and political affiliation (such as whether they live in a top 10 red city or top 10 blue city) on the following two metrics: (i) social distancing and (ii) privacy behavior. We focused on demographics, because a number of articles from both academia and the mainstream media have discussed heterogeneity in behavior as driven by the above-noted demographic factors. Prior research has shown that the compliance rates of non-therapeutic interventions among young adults are less than the average compliance rates, because young adults in the U.S. do not follow protective behaviors to the same extent as other segments of the population do.[15] Lower compliance among them makes it easier to spread infections, as they have higher levels of human contact. Specifically, we focused on three main research questions: (1) how overall privacy choice (i.e., opt out) respecting mobile location-data sharing has changed before and after COVID-19 being declared a national emergency, (2) how such changes in privacy choice vary across demographic groups in blue (Democratic) cities and red (Republican) cities, and (3) what the relationship between individuals' practice of social distancing and willingness to share mobile location data during COVID-19 is, and how it varies with political affiliation. Our analyses revealed three major findings. First, there is a significant decreasing trend of mobile-app location-sharing opt out in the U.S. Whereas areas with more Democrats were more privacy-concerned than areas with more Republicans before the advent of the COVID-19 crisis, there was a significant decrease in overall opt-out rates after COVID-19, which effect was more salient among Democratic than Republican cities. Second, people who practice social distancing (i.e., those who travel less and interact with fewer close contacts during the pandemic) are less likely to opt out, while the converse is true for those less compliant with social distancing. This effect appeared to be more salient among Democratic than Republican cities after COVID-19. Third, high-income people and males are more privacy-concerned, and thus more likely to opt out of location tracking. This research endeavored to make a valuable contribution to the literature of privacy and pro-social consumer behaviors by demonstrating a powerful pro-sociality-motivated relinquishing of personal privacy, particularly as massive granular individual-level location data have become an increasingly prominent tool for academic research, policy making, and combating the present, unprecedented pandemic.

## 2. Literature

Our research is closely related to the literature on consumer privacy and pro-social behaviors as well as social distancing, political affiliation and demographics amid the COVID-19 pandemic. Below we will concisely review each literature.

---

[15] Singh et al. BMC Infectious Diseases (2019) 19:221. https://doi.org/10.1186/s12879-019-3703-2.

## 2.1 Consumer Privacy and Pro-social Behavior

Granular location data and other consumer data have generated widespread concern for consumer privacy (Wedel and Kannan 2016). In the business context, privacy broadly pertains to the protection of individually identifiable information online and offline along with the adoption and implementation of privacy policies and regulations. In the context of location data, privacy concerns may arise from, for instance, identification of sensitive information such as home address, workplace location, daily movements, or social-distancing compliance. There is a rich literature on Information Systems and Marketing as it pertains to consumer privacy, the data sources being surveys (Acquisti et al. 2012), direct marketing (Goh et al. 2015), and digital advertising (Goldfarb and Tucker 2011; Ghose 2017). Researchers also have investigated how consumers make privacy choices of platform-provided privacy settings (e.g., Burtch et al. 2015; Adjerid et al. 2019) and how they select opt-in/out options provided by email marketing programs (Kumar et al.2014). Overall, this research points to the positive effects of granting consumers enhanced control over their own privacy.

Besides egocentric motivations such as financial benefits (Soleymanian et al. 2017), consumers may choose to relinquish privacy for altruistic reasons, such as to support national security post-9/11 in the form of the Patriot Act and its broad expansion of the federal government's surveillance powers.[16] Similarly, amid an unprecedented pandemic like COVID-19, some people may reduce their opt out of location tracking for the social good, such as for digital contact tracing and city-wide social-distancing analyses to identify clusters of outbreak and hotspots. Although the specific research on the potential tradeoff between privacy concerns and pro-social behaviors is limited, studies such as that by Burtch, Ghose, and Wattal (2015) in the context of online fund-raising campaigns for charities have shown that reduced privacy settings (e.g., access to information controls for concealment of personal identity or contribution amount) increases pro-social behaviors. Whereas monitoring such as through location tracking and contact tracing can cause uneasiness, the feeling of being observed and accountable can incentivize people to engage in pro-social behaviors or adhere to social norms (Acquisti et al. 2015).

The literature on pro-social consumer behavior has also demonstrated various motivations for consumers' pro-social behaviors, including both egocentric motivations (e.g., ego utility, self-signaling, and reputation benefits) and altruistic motives (Gershon et al. 2020). Related to the present COVID-19 pandemic, sympathy biases, such as for identifiable victims, are also shown to drive the extent of pro-social behaviors (Sudhir et al. 2016). As demonstrated by this literature, pro-social tendency can be escalated by contextual factors. Even general coverage of climate change or global warming by major media outlets, for instance, can exert an overall positive impact on the sales of hybrid vehicles (Chen et al. 2019). In times of sudden and widespread crises such as the COVID-19 pandemic, filled as they are with life-and-death decisions, negative emotions, as well as divergent risk perceptions and behavioral norms, it is plausible that consumers would exhibit heightened pro-social behaviors as exemplified by better compliance with shelter-at-home policies to protect others' health and reduced opt outs in location tracking from mobile devices to support various initiatives, such as social-distancing monitoring, contact tracing, and other public safety measures. Bagozzi and Moore (1994) show that public service ads designed to reduce the incidence of child abuse stimulate negative emotions that result in empathic reactions and increased willingness to help.

---

[16] After 9/11, we gave up privacy for security. Will we make the same trade-off after COVID-19? https://www.statnews.com/2020/04/08/coronavirus-will-we-give-up-privacy-for-security/.

Finally, there is also evidence that most people do not make the effort to truly understand the privacy policies of mobile apps or websites, primarily because they are starved for time. A 2008 study estimated that it would take 244 hours a year for the typical internet user to read the privacy policies of all websites he or she visits— and that was before everyone carried smartphones with dozens of apps.[17] Another analysis, this one on the length and readability of privacy policies based on nearly 150 popular websites and apps, shows that it can take up to 18 minutes to read the privacy policies on certain platforms.[18] Schaub et al. (2017) argue that today's privacy notices and controls are surprisingly ineffective at informing users or allowing them to express choice. MacDonald and Cranor (2008) demonstrate that it takes about 8 to 12 minutes to read privacy policies on the most popular sites, their point estimate being 10 minutes per policy. With the increased time available due to shelter-at-home and social-distancing policies, at least for certain demographics and societal segments, it is conceivable that some consumers may finally have the time to read through the mobile apps' privacy policies and react (either change their choices or maintain the status quo) accordingly given heightened surveillance concerns and awareness.

**2.2 Social Distancing, Political Affiliation, and Demographics**

Amid the COVID-19 pandemic, social distancing has resulted in empirically verified, dramatic changes in where people spend their time and with how many people they interact. For instance, based on an analysis of anonymized location data in New York City, Bakker et al. (2020) show that distance travelled everyday dropped by 70%; the number of social contacts in certain places decreased by 93%, and the number of people staying home the whole day increased from 20 to 60%. Measuring the relative transmission-risk benefit and social cost of closing about thirty different location categories in the U.S. across eight dimensions of risk and importance and through composite indexes, Benzell et al. (2020) find that from February to March, there were larger declines in visits to locations that their measures imply should be closed first. Overall, while social distancing may reduce the population livelihood and entail personal inconvenience, it is shown to lead to substantial economic benefits. Greenstone and Nigam (2020) estimate that moderate distancing beginning in late March 2020 would save 1.7 million lives by October 1, and that the mortality benefits of social distancing would be about $8 trillion or $60,000 per household. Social distancing is thus widely perceived, amid the COVID-19 pandemic, as a pro-social behavior.

---

[17] https://www.salon.com/2017/10/14/nobody-reads-privacy-policies-heres-how-to-fix-that_partner/
[18] https://www.nytimes.com/interactive/2019/06/12/opinion/facebook-google-privacy-policies.html

A number of recent studies have demonstrated a strong link between social-distancing compliance and political affiliation. For example, a higher percentage of Republican county vote share was found, based on location data, to be associated with increased social-distancing non-compliance, even after the declaration of a national emergency, and that, according to survey data, Democrats believe the pandemic to be more severe and report a greater reduction in contact with others (Fan et al. 2020). Also, based on location data, areas with more Republicans engage in less social distancing, and according to survey data, significant gaps exist between Republicans and Democrats in beliefs about personal risk and the future path of the pandemic (Allcott et al. 2020). Similarly, using internet search and location data, Barrios and Hochberg (2020) reveal that a higher percentage of Trump voters in a county is associated with a lower perceived risk associated with COVID-19 and less social distancing.

Other studies have revealed a strong link between social-distancing compliance and demographics such as gender, income, and race. For instance, low-income, black, and Hispanic neighborhoods in New York City exhibit more work activity during the day and less sheltering in place during non-work hours (Coven and Gupta 2020). Similarly, Ruiz-Euler et al. (2020) present widespread evidence of a mobility gap, i.e., the decline in human mobility during COVID-19 lockdown happened at different speeds for high versus low income groups within most cities, as lockdown imposes low-income groups with a stringent choice between health and income. Wright et al. (2020) further suggest that poverty reduces social-distancing compliance. Conversely, Chiou and Tucker (2020) demonstrate that people from regions with either high-income or high-speed Internet display more compliance with state-level directives to remain at home, which finding suggests an impact of the digital divide on the ability to comply with social-distancing policies. Painter and Qiu (2020) show that Democrats are less likely to respond to a state-level order when it is issued by a Republican governor relative to one issued by a Democratic governor.

Our study distinguishes itself from the extant literature in the following ways: (1) it is based on unique, massive, *individual-level* location data, whereas nearly all of the prior work leveraged *aggregate* location data; (2) it examines opt-out decrease as a valuable pro-social indicator of the criticality of location tracking in the unprecedented combat against the COVID-19 pandemic.

## 3. Data

For the purposes of our analysis, we combined two datasets: individual-level GPS location tracking data, and census-block-level demographic data from The American Community Survey (2016).[19] For the location data, we collaborated with a leading data collector that aggregates location data across hundreds of commonly used mobile applications ranging from news to weather, map navigation, and fitness. The location-data collection was performed through a GDPR- and CCPA-compliant framework. The data covers one-quarter of the U.S. population across the Android and iOS operating systems.[20] Each row of the data corresponds to a location recorded for an individual. Each row contains information about 1) *User ID* - an anonymized unique identifier of an individual using a mobile app, 2) *Latitude*, *longitude* and *timestamp* of a location visited, 3) *Speed* at which the location was captured, and 4) *App category* - the type of mobile app, such as news or weather, that captured the location.

In total, we collected, from January 1 to April 15th, 2020, detailed, fine-grained location data on individuals in 20 major U.S. cities including Baltimore (MD), Washington D.C., Boston (MA), San Francisco (CA), New Orleans (LA), New York City (NY), Seattle (WA), Pittsburgh (PA), Philadelphia (PA), Austin (TX), Phoenix (AZ), Arlington (TX), Oklahoma City (OK), Wichita (KS), Nashville (TN), Omaha (NE), Lexington (KY), Colorado Springs (CO), Virginia Beach (VA), and Jacksonville (FL).[21] We choose these cities based on the respective political affiliations shown in recent national elections. Among the 20 cities, the former ten are rated as among the Top-10 most liberal cities, and the latter ten, the Top-10 most conservative cities.[22] For each of the 20 cities, we parsed through an average of 150k individuals, 70 locations per individual per day, 1.5M unique locations overall, and 1.1B rows of individual location data.[23] To create the final panel dataset for our empirical analysis, we selected a random sample of 25,000 individuals[24] per city. For each individual, we identified the home address[25] and assigned the demographics of the census block that was closest[26] to the home address.

---

[19] We obtained Census Block Group data from SafeGraph: https://docs.safegraph.com/docs/open-census-data#section-census-demographic-data.

[20] Although due to data anonymity we did not observe the demographics of each individual user in our data, we did observe the geographical distribution of users' home locations covering all census blocks in each city. In addition, we also observed users who opted out of location tracking for nearly all census blocks in a given city (instead of a selected subset of blocks). Put simply, there was no systematic selection for opt-outs. Therefore, our user sample can be considered to be well representative of the U.S. populations in the cities studied.

[21] Each city is defined as including the Metropolitan Statistical Area.

[22] Tausanovitch, C. and Warshaw, C. 2014. "Representation in Municipal Government," *American Political Science Review* (108:3), pp.605-641. https://ctausanovitch.com/Municipal_Representation_140502.pdf.

[23] Depending on the app setting, the locations are typically tracked every 5-20 minutes throughout the day, or when a person moves out beyond 100 meters even within 5 minutes.

[24] We tried different sizes of random samples and found 25k to be a rather representative sample with a highly consistent data distribution as the full sample.

[25] We assigned the most frequent location captured during 3 - 5 AM to an individual as their home address. A similar heuristic was used to identify home address from location data in a previous study (Macha et. al 2019).

[26] For assignment of the census block, the Haversine distance between the individual's home address and the interpolated center latitude and longitude of the Census Block Group provided by SafeGraph was computed.

## Table 1a. Variable Description

| Variable Set | Level | Variable Name | Variable Description |
|---|---|---|---|
| Privacy Outcome | Individual | OptOut | Indicator of individual who stopped sharing location data - opted out of location tracking |
|  | Block | OptOutCount | Number of people who stopped sharing location data on certain day (opted out) |
| Privacy Controls | Block | TotalActiveUsers | Total daily number of people sharing location data (opted in) |
|  | Block | TotalNewUsers | Number of people who newly started sharing location data in block. |
| Social Distancing | Individual /Block | TotalContacts | Sum of daily unique contacts for all individuals |
|  | Individual /Block | TotalDistance | Average of daily total distance traveled (km) |
|  | Individual /Block | AvgSpeed | Average of daily mean speed (kmph) |
| COVID-19 Health Risk | City/County | Infection Rate | Ratio of daily infected COVID-19 cases by population |
|  | City/County | Death Rate | Ratio of daily deaths by infected cases |
| Social Demographics | Block | Gender | Proportions of male and female populations in each block |
|  | Block | Income | Proportions of households with income <60K, 60-100K, 100-150K, 150-200K, >200K |
|  | Block | Race | Proportions of White, Black, Asian, and Native Indian populations in each block |
|  | Block | Population Density | Population Amount of land (in acres) in each block |
| App Usage | Individual /Block | TotalAppUsage | Total duration of mobile app use on specific day (minutes) |
|  | Individual /Block | AppUsageCategory | Number of unique mobile app categories that people use on specific day |
|  | Individual /Block | AppUsageHHI | Herfindahl–Hirschman Index (HHI) defined using mobile app use duration by categories |
| Political Affiliation | City | 5 "Blue" Cities 5 "Red" Cities | Designated based on previous national elections Baltimore, DC, Boston, SFO, New Orleans Omaha, Lexington, Colorado Springs, Virginia Beach, Jacksonville |
| Treatment | Country | Treat | Whether it is after declaration of National Emergency (on or after Mar. 13, 2020) |

**Table 1b. Variable Summary Statistics**

| Variable Set | Variable Name | Mean | Std. Dev. | Min | Max |
|---|---|---|---|---|---|
| Privacy Outcome | OptOut | 0.02 | 0.005 | 0 | 1 |
| | OptOutCount | 0.32 | 1.41 | 0 | 219 |
| Privacy Controls | TotalActiveUsers | 19133 | 5074 | 5714 | 24942 |
| | TotalNewUsers | 0.35 | 1.56 | 0 | 1900 |
| Social Distancing | TotalContacts | 5.59 | 23.78 | 0 | 787 |
| | TotalDistance | 6.56 | 7.74 | 0 | 423 |
| | AvgSpeed | 3.87 | 4.60 | 0 | 129 |
| COVID-19 Health Risk | Infection Rate (%) | 0.04 | 0.12 | 0 | 1.02 |
| | Death Rate (%) | 0.01 | 0.03 | 0 | 0.36 |
| Social Demographics | Gender (Male) | 0.48 | 0.08 | 0.10 | 0.94 |
| | Gender (Female) | 0.51 | 0.08 | 0.06 | 0.90 |
| | Income (<60K) | 0.54 | 0.24 | 0 | 1 |
| | Income (60-100K) | 0.20 | 0.11 | 0 | 1 |
| | Income (100-150K) | 0.14 | 0.10 | 0 | 1 |
| | Income (150-200K) | 0.07 | 0.07 | 0 | 1 |
| | Income (>200K) | 0.09 | 0.12 | 0 | 1 |
| | Race (White) | 0.55 | 0.31 | 0 | 1 |
| | Race (Black) | 0.27 | 0.32 | 0 | 1 |
| | Race (Asian) | 0.09 | 0.14 | 0 | 1 |
| | Race (Native Indian) | 0.00 | 0.01 | 0 | 0.28 |
| | Population | 1339 | 840 | 3 | 15096 |
| | Amount of land (in acres) in each block | 372 | 1858 | 3.34 | 50569 |
| App Usage | TotalAppUsage | 5.13 | 4.46 | 0 | 53.86 |
| | AppUsageCategory | 0.28 | 0.21 | 0 | 5 |
| | AppUsageHHI | 0.28 | 0.2 | 0 | 1 |
| Total # Observations: | 22 Billion | | | | |
| Observation Period: | January 1st, 2020 - April 15th, 2020 | | | | |

In Tables 1a and 1b, we detail and provide the summary statistics (across all cities) for the four types of measures computed from the location and census block data.

1. *Privacy Choices*: At the day level, an individual was considered "opt in" for location tracking on all days between the first and the last day we observed the individual's locations in our data. If we never saw an individual's records in the data after a particular day, we considered that individual to be "opt out" after that last day (*OptOut*).[27] [28] We computed *OptOutCount* at the block-day level as the total number of individuals who opted out on that day in a given block. These two measures were our primary dependent variables modeled in our empirical analysis.[29] For controls, we computed the total number of current opt-in individuals on a given day in each city (*TotalActiveUsers*), as well as the total number of new opt-in individuals first observed on that day and block (*TotalNewUsers*).

2. *Social Distancing*: We determined "contact" between two individuals if they were co-located in the same 10-meter radius at a popular location[30] within a 15-minute window. This was motivated by various COVID-19 media studies that show that human-to-human transmission can occur when individuals come in close contact within similar distance and time windows (e.g., Benzell et al. 2020).[31] At the block level, we summed the daily unique contacts for all individuals in the block (*TotalContacts*). To estimate the distance traveled by an individual, we computed the Haversine distance between two consecutive stay locations (*TotalDistance*)[32].

3. *COVID-19 Health Risk*: We considered two major measures of COVID-19 health risk: *InfectionRate*, which measures the ratio of the daily number of infected cases to the total population, and *DeathRate*, which measures the ratio of the daily number of deaths to the total number of infected cases.

4. *App Usage*: For each individual, we derived the mobile app usage behavior every day using the timestamp and category information. Our data vendor provided a list consisting of approximately 30 app categories[33]. For each individual each day, we calculated the total app usage duration (*TotalAppUsage*), the number of unique app categories used (*AppUsageCategory*), and the Herfindahl–Hirschman Index

---

[27] There is a chance when an individual uninstalls an app (instead of opting out of location sharing) that it will also lead to the cessation of location data. However, to completely disappear from our data, the individual needs to simultaneously uninstall all of the (hundreds of) affiliated apps that use the location-tracking SDK provided by our location data collector. Given the market share of the company and the coverage of their client apps, this was very unlikely to occur. In addition, we also conducted robustness tests to check individuals' app usage and location data points tracked before the last day of observation (i.e., before the user opted out). If our result were due to uninstallations of all of the affiliated apps over time, we would expect to see an overall decreasing trend in daily app usage time, unique app categories, and the daily number of locations tracked over time before the opt-out. We did not find any such significant decreasing trend. We provide the detailed analyses in Section 6.3.

[28] It is also possible that an individual opts out of location tracking and then decides to opt in again, and then opts out again, etc. Based on our conversation with the company, very few users demonstrate such behavior. For our analysis, we adopted a conservative definition and considered only the last opt-out of the same user. In addition, note that when a user decides to opt out of location tracking, s/he can opt out of a specific app (each app has an opt-out option), but can also choose to opt out of all apps in our partner company's affiliated app network. In our study, we defined opt out as the strictest case wherein the user opts out of all apps. Therefore, our measure is a lower bound of the actual opt-out rate.

[29] For robustness checking, we also considered "opt in" instead of "opt out" (the user opt in and block-day-level total opt-in counts) as alternative outcome variables, and the results were highly consistent.

[30] Top 1% of locations based on popularity - number of unique individual visits to the locations. We also performed a sensitivity analysis based on alternative definitions of "contact" by varying the notion of location popularity (1.5 and 2%), and noticed that the distribution of the number of unique contacts remained similar.

[31] https://www.cdc.gov/coronavirus/2019-ncov/prevent-getting-sick/how-covid-spreads.html;
https://www.thelocal.de/20200417/what-you-need-to-know-about-plans-for-germanys-states-to-ease-lockdown.

[32] To compute the distance, we considered only locations more than 10 meters apart and where individuals have spent at least 1 minute each.

[33] We observed major app categories including weather, news, fitness, health, education, business, social networking, etc.

(HHI) determined using category-level duration (*AppUsageHHI*) that captures the concentrations of preferences.

5. <u>S*ocial Demographics*</u>: In addition to the individual-/block-level measures from the location data, we also compiled block-level social demographics (population density, proportions of different races, income levels, and gender).

6. *Treatment Indicator*: We defined treatment as the declaration of the National Emergency for COVID-19 by President Trump on March 13, 2020. This definition was motivated by the fact that during the same time period (March 10-20), news headlines appeared in mainstream media such as CNN[34] and Fox News[35] leading many people to become aware of contact tracing, location-data tracking, and potential surveillance concerns.

Figure 1 shows the detailed, anonymized location data from twenty randomly sampled individuals with 2318 (2465) unique (total) locations in SFO on a certain day in the January 1-April 15 period.

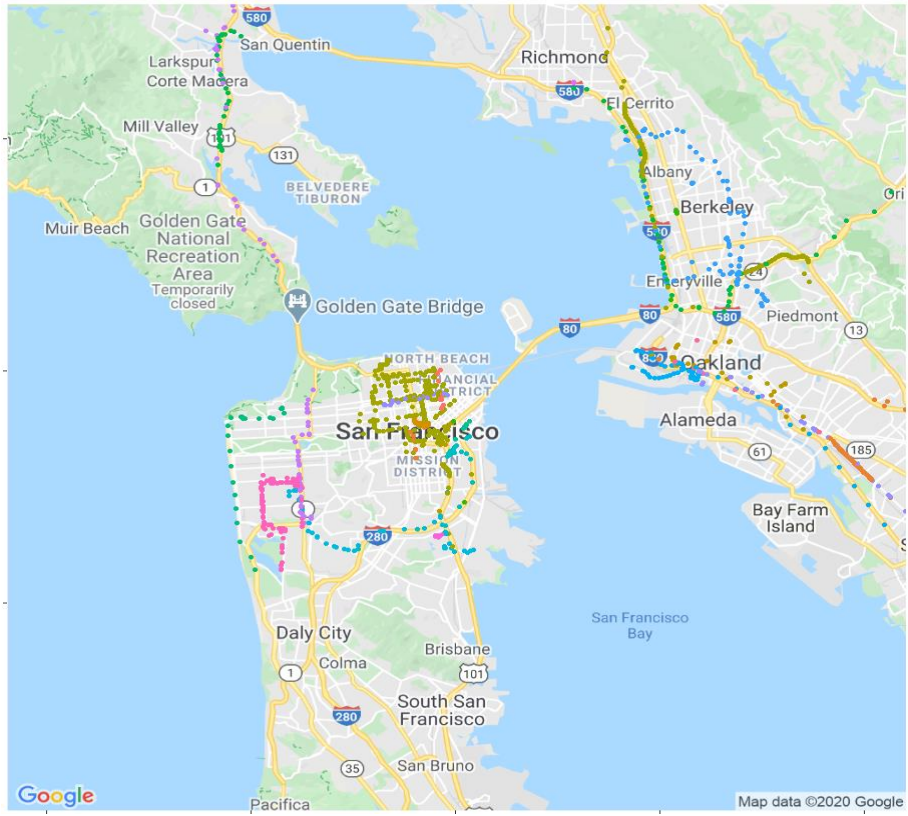

**Figure 1. Sample GPS location data from 20 randomly sampled individuals in SFO**

---

[34]https://www.cnn.com/2020/03/10/health/coronavirus-contact-tracing/index.html; https:/3/www.cnn.com/2020/03/18/tech/us-government-location-data-coronavirus/index.html.
[35]https://www.foxnews.com/tech/us-government-big-tech-smartphone-coronavirus-google-facebook; https://www.foxnews.com/tech/taiwans-so-called-electronic-fence-monitor-for-those-quarantined-raises-privacy-concerns-report.

## 4. Model-Free Evidence

Before specifying the econometric model to quantitatively understand privacy choices, we present qualitative model-free evidence. In Figure 2, we first show the overall distribution of monthly total opt outs across all 20 cities, one month before and one month after the national emergency declaration regarding COVID-19 on March 13. As we can see, there appeared to be a consistent decreasing trend in the opt outs in each city after COVID-19. To further validate this finding, we next plot detailed day-level trends.

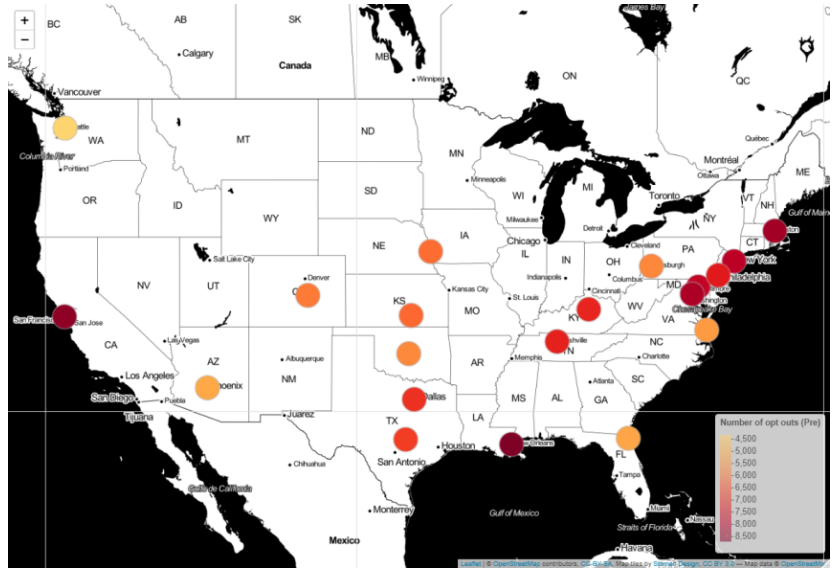

(2a) One Month before Mar. 13

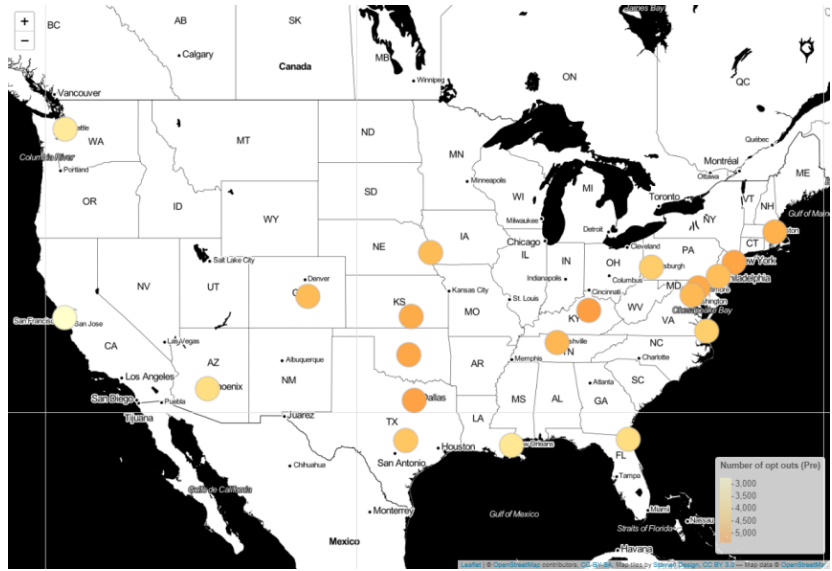

(2b) One Month after Mar. 13

**Figure 2. Total monthly opt outs in each of 20 cities one month (a) before / (b) after COVID-19 national emergency (darker-red/light-yellow color representing more/fewer opt outs)/**

More specifically, we show the model-free evidence for our three major findings in Figures 3-7. In each of the figures, we plot day-level trends of average opt-out rates across the January 1 - April 7th, 2020 period for a different grouping of blocks. The dotted line in each figure indicates March 13, the day President Trump declared the national emergency.

## 4.1 Political Affiliation and Privacy Concern

In Figure 3, we plot the day-level opt-out rates by pooling the 10 red and blue cities (and their corresponding blocks), respectively, together. From the figure, we observe that in general, before the COVID-19 emergency declaration, blue cities were more privacy-concerned than red ones. This is consistent with recent studies showing that Republicans are less concerned about privacy than Democrats.[36] Interestingly, we see that both red and blue cities show a decreasing trend in opt out after the declaration. Prominently, the decrease in opt-out rates is more salient in blue cities than in red ones. In addition, we also notice an interesting weekend/weekday effect—people are more likely to opt out during weekends than weekdays. This pattern becomes less prominent after COVID-19, potentially due to the blurred boundary between weekends and weekdays after the stay-at-home order.

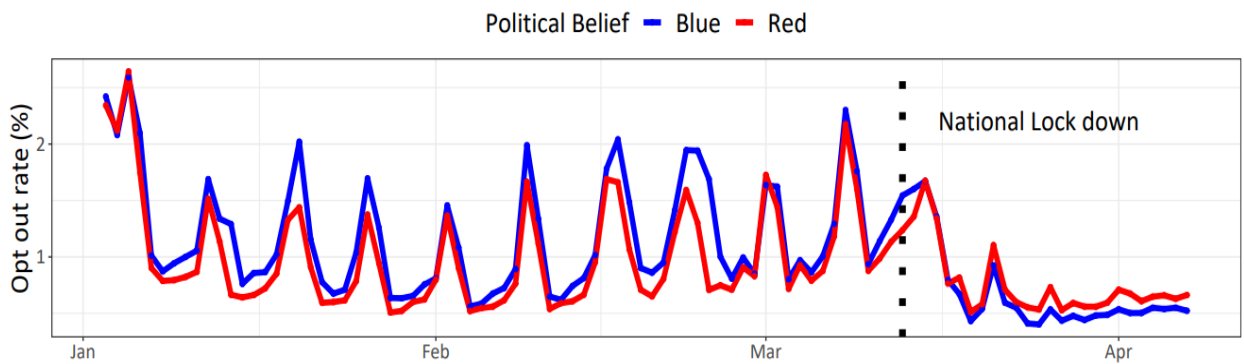

**Figure 3. Day trends of opt-out rates for red and blue cities**

## 4.2 Social Distancing and Privacy Concern

In Figure 4, we plot the day-level opt-out rates for the 20 cities, where the blocks in each city are grouped by the number-of-contact quantiles. From the plots, we observe that in general, blocks with lesser social-distancing practices (higher number-of-contact quantiles) have lower willingness to share location data (higher opt-out rates) compared with blocks with better social-distancing practice. We also observe that in

---

[36] "Divided We Feel: Partisan Politics Drive Americans' Emotions Regarding Surveillance of Low-Income Populations," https://www.nytimes.com/2018/04/30/technology/privacy-concerns-politics.html.

Virginia Beach and San Francisco, the trend of the opt-out rate stays relatively the same for people practicing lesser social distancing even after the declaration.

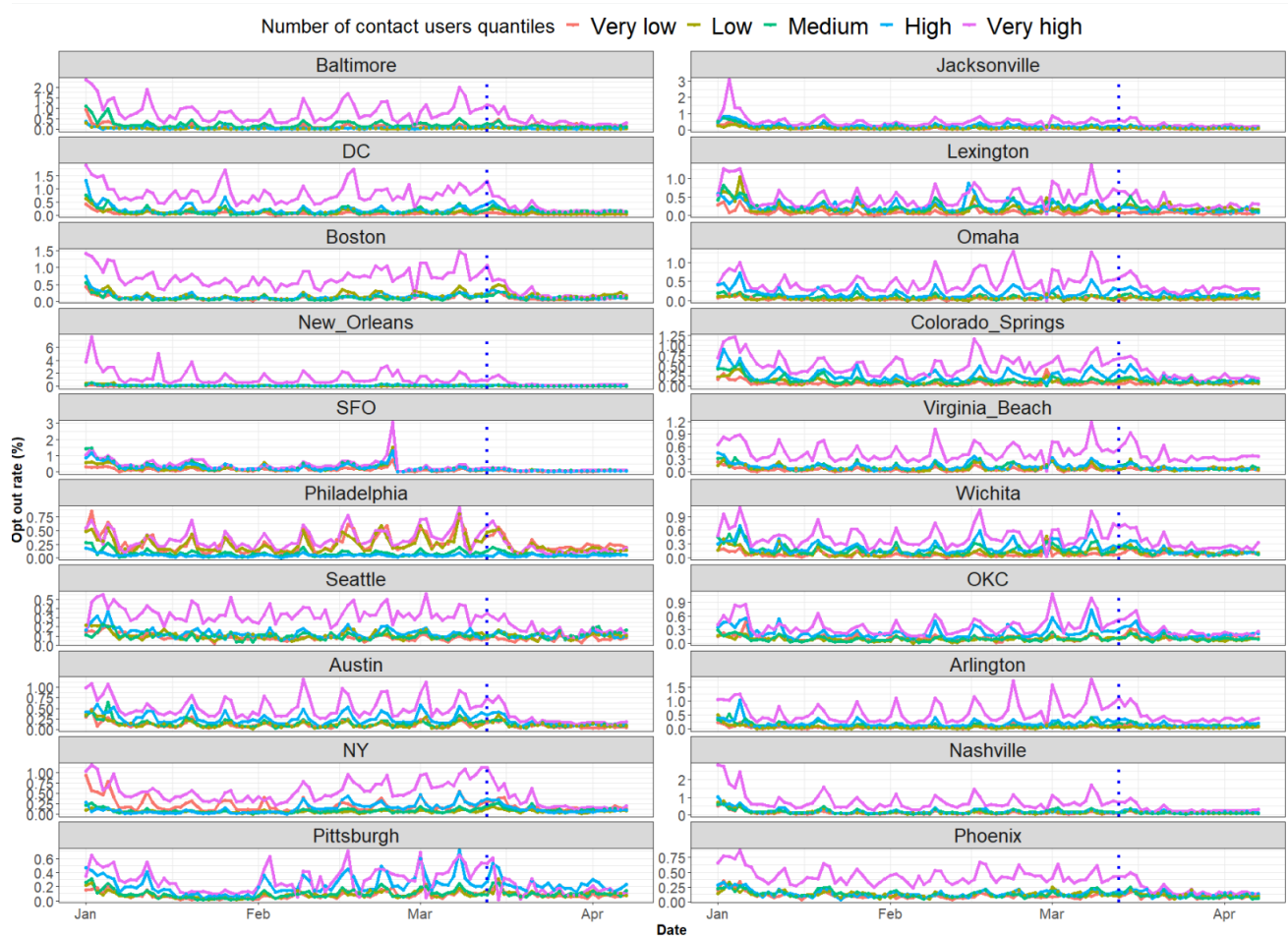

**Figure 4. Day trends of city-level opt-out rates for different number-of-contact quantiles (Left, Blue cities; Right, Red cities)**

### 4.3 Social Demographics and Privacy Concern

In Figure 5, we plot the opt-out rates for different cities, where the blocks are grouped based on the proportion of households with income more than $200K. Hence, blocks with a lower/higher proportion of households with income more than $200K belong to low-/high-income quantiles. In Figure 5, we see that in many cities (e.g., Boston, New Orleans, Philadelphia, NYC, Pittsburgh, Lexington, Wichita) high-income people are more privacy-concerned compared with low-income people. After the emergency declaration, the drop in opt-out rates seemingly is heterogeneous across the different income groups.

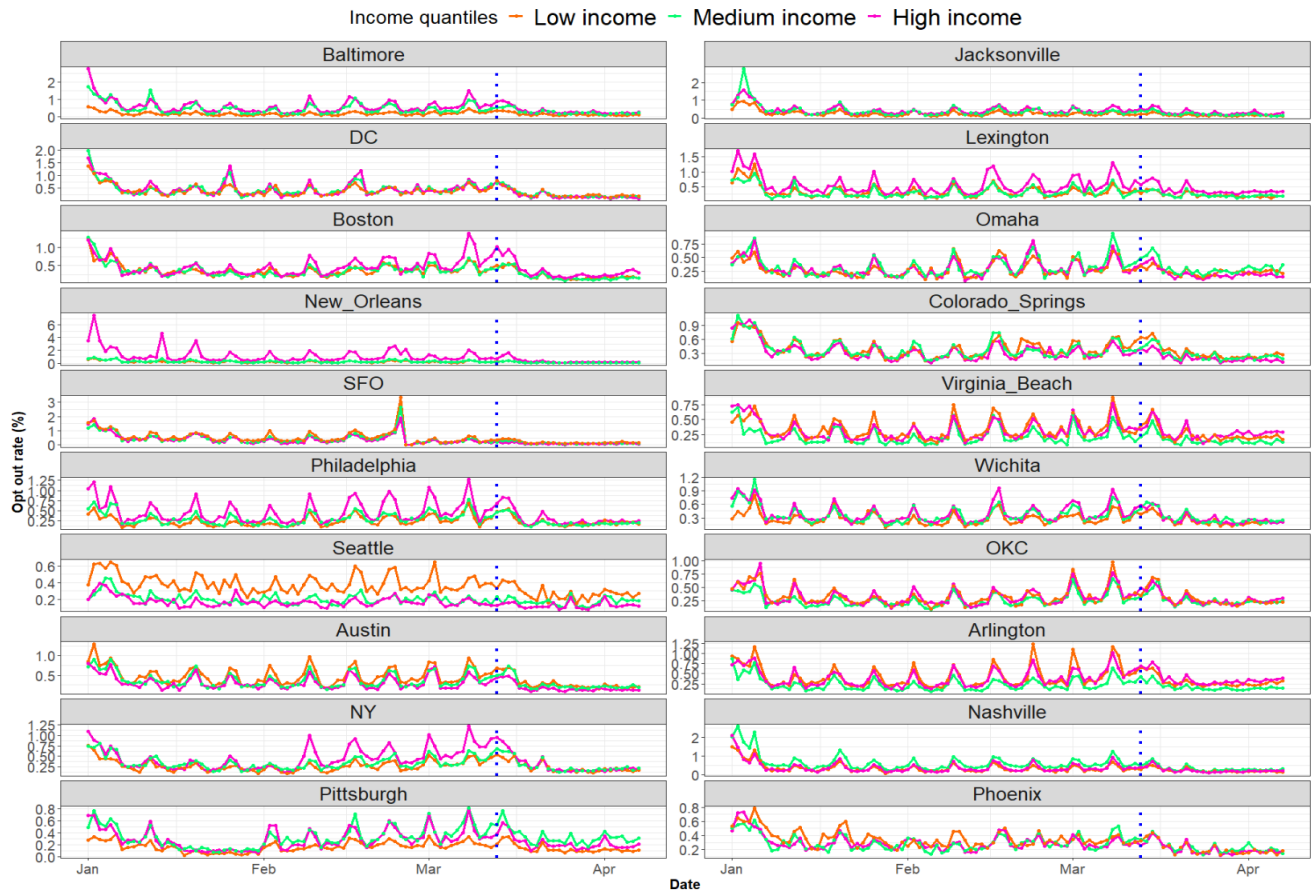

**Figure 5. Day trends of city-level opt-out rates for different income quantiles**
**(Left, blue cities; Right, red cities; three quantiles – from highest to lowest income)**

In Figure 6, we compare the opt-out rates across gender by grouping blocks with a higher/lower proportion of males. We observe that blocks with a higher male population are in general more privacy-conscious than their counterparts. We can see that after the declaration, there is an overall drop in opt-out rates across all cities and across both genders.

Finally, in Figure 7, we plot the opt-out rates across different racial diversities by grouping blocks with a higher/lower proportion of the white population. In Baltimore, DC, and New Orleans (all blue cities), we observe that blocks with a higher white population have a higher opt-out rate relative to more diverse census blocks. In Boston (blue), Jacksonville, and Omaha (red), we notice that people in relatively lower-white-populated blocks (2nd quantile in Figure 7) tend to opt out more than do others. In all of Figures 4-7, we notice that San Francisco shows little heterogeneity across all social demographics.

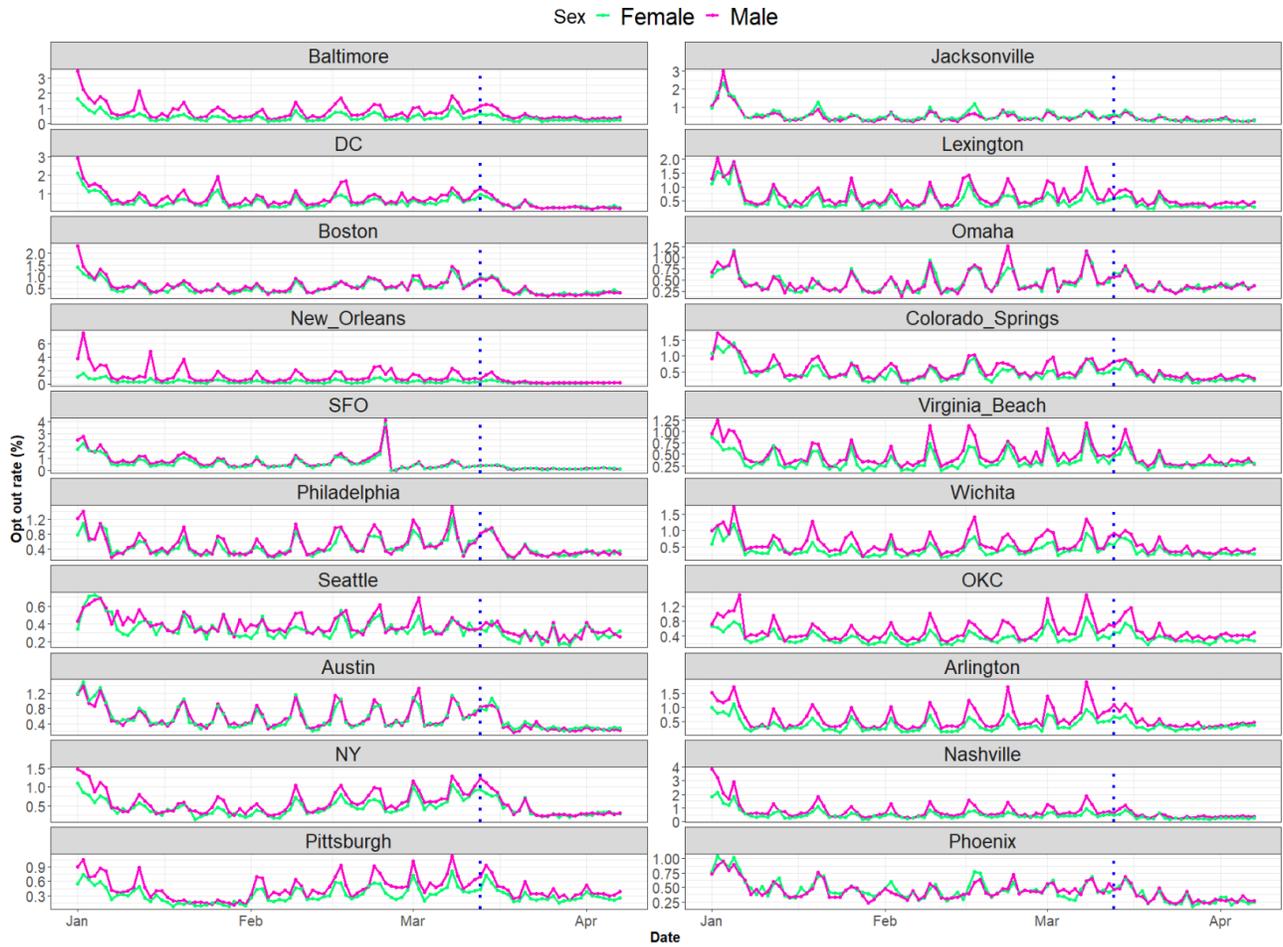

**Figure 6. Day trends of city-level opt-out rates for males and females (Left, blue cities; Right, red cities)**

## 5. Empirical Methodology

While the qualitative trend insights regarding opt-out rates from Figures 2-7 are useful, they do not control for many other shifts that were taking place during the time period. Therefore, to further examine the effects of social demographics, social distancing, and political affiliation on American's privacy choices during COVID-19, we conducted detailed econometric analyses using panel data models at both the individual-day level and the census-block-day level in each of the 20 cities.

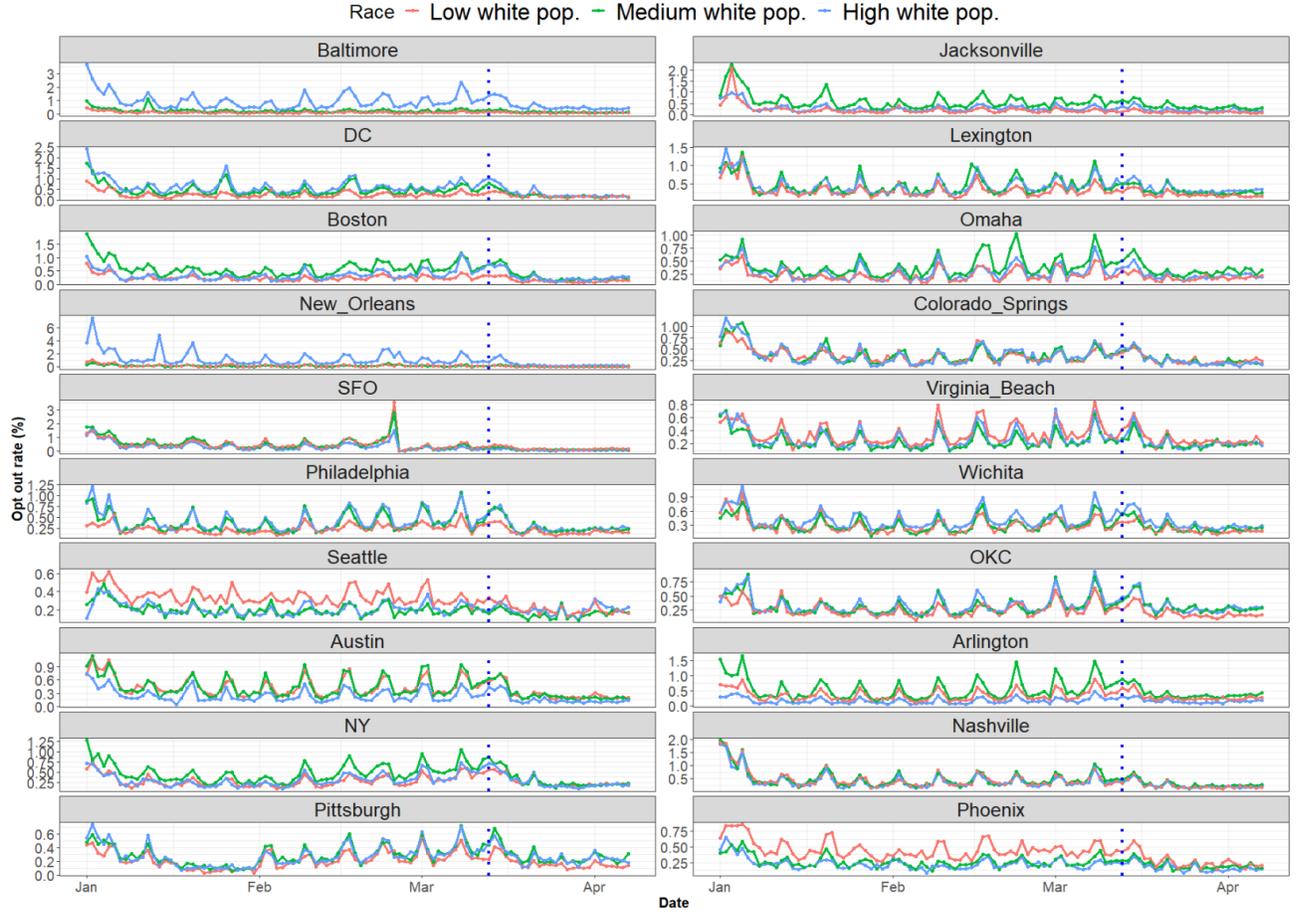

**Figure 7. Day trends of city-level opt-out rates for different racial diversities (Left, blue cities; Right, red cities)**

## 5.1 Social-Demographic Analyses

More specifically, first we analyzed the effect of social demographics on an individual's likelihood of opting out of mobile location tracking before and after President Trump declared the national emergency on March 13, 2020.[37] We modeled the likelihood of the individual's opt-out choice using the logit model

$$\Pr(OptOut_{it}) = \frac{\exp(\overline{U_{it}})}{1+\exp(\overline{U_{it}})}, \qquad [1]$$

where

$$\overline{U_{it}} = \alpha_{it} + \beta_1 Treat_t + \beta_2 Treat_t * D_i + \beta_3 D_i + \beta_4 Week_t + \beta_5 DayOfWeek_t +$$
$$\beta_6 AppUsageControl_{it} + \beta_7 TotalActiveUsers_t + \beta_8 TotalNewUsers_t + \beta_9 OtherControl_i + \beta_{10} T_t$$

$$[2]$$

---

[37] We also examined, as an alternative, the treatment effect of the state's specific lockdown effects and found the results to be qualitatively similar. We will discuss this in more detail in Section 6, Table 5.

$i$ represents each individual and $t$ represents each day; $Treat_t$ is the binary treatment variable that indicates whether or not it was after the declaration of the national emergency, and $D_i$ are vectors containing the corresponding income/gender/race groups. For our analyses, we considered five different income groups (<60K, 60K-100K, 100K-150K, 150K-200K, >200K), two gender groups (Male, Female), and five racial groups (White, Black, Asian, Native Indian, Others). The interaction effect $Treat_t * D_i$ is our variable of interest that captures the heterogeneous treatment effect of the COVID-19 national emergency declaration across different social-demographic groups.[38]

$Week_t$ and $DayOfWeek_t$ are dummies to control for the week and day-of-week fixed effects. $T_t$ is a day index (from 1 to 106) to control for any continuous time trend from January 1 to April 15, 2020. $AppUsageControl_{it}$ controls for all of the user-day-level app usage factors, such as the user's total app usage time, the total number of app activities, the total number of unique app categories, and the entropy of usage time across the different app categories on a certain day. Moreover, we considered $TotalActiveUsers_t$ and $TotalNewUsers_t$ to further control for the daily-level trend of user opt in to mobile location tracking. We ran the individual logit model for each of the 20 cities. In addition, $OtherControl_i$ captured any other additional factors such as the population density of the block where the individual resides, as well as the COVID-19-related health risk (infection rate and death rate).

For robustness checking, we also ran these analyses using individual-level and day-level fixed effects. Notably, nonlinear models (e.g., Logit) sometimes are difficult to interpret, because the cross-partial may have a different sign than the coefficient on the interaction term (Ai and Norton 2003). So, to check the robustness of the coefficients in the interaction terms in the Logit model, we also conducted a similar analysis using the Linear Probability Model (LPM).[39] Overall, we found that the interaction effects remained qualitatively consistent. Our main results are shown in Table 2. We will discuss them in the next section.

## 5.2 Social-Distancing and Political-Affiliation Analyses

Beyond the social-demographic effect, we were also interested in the relationships among social distancing, political affiliation, and Americans' privacy choices before and after COVID-19. To study such relationships, we first examined the relationship between American's daily social-distancing practice and the daily number of people choosing to opt out of location tracking at a census block level in each of the 20 cities. For each city, we conducted a panel data analysis using the Poisson model:

$$E(OptOutCount_{jt}) = \exp(\alpha_{jt} + \beta_1 Treat_t + \beta_2 Treat_t * SocialDistancing_{jt} + \beta_3 SocialDistancing_{jt} + \beta_4 Week_t + \beta_5 DayOfWeek_t + \beta_6 AppUsageControl_{jt} + \beta_7 TotalActiveUsers_t + \beta_8 TotalNewUsers_t + \beta_9 OtherControl_j + +\beta_{10} T_t + \varepsilon_{jt}),  \quad [3]$$

where $j$ represents each census block, and $t$ represents each day; $SocialDistancing_{jt}$ contains an individual's daily social-distancing measures aggregated at the block level, such as the daily total number of close contacts, the daily total traveling distance, the daily average traveling speed, the daily total number of unique locations visited, and the daily average time spent at each different location; $Treat_t$ is the treatment indicator for whether or not it was after President Trump's declaration of the national emergency, and the interaction effect $Treat_t * SocialDistancing_{jt}$ captures the heterogeneous treatment effect of the COVID-19 national emergency declaration across different social-distancing groups. We also ran the same model without the interaction effect to examine the mean effect of COVID-19 on Americans' privacy choice, as well as the relationship between people's practice of social distancing and willingness to share their location data. Similarly to Equation [2], we also controlled for all other factors including $Week_t$ and $DayOfWeek_t$ dummies, $T_t$ time trend, $AppUsageControl_{jt}$, $TotalActiveUsers_t$, $TotalNewUsers_t$, and $OtherControl_j$.

---

[38] To be precise, the mean treatment effect estimated from our model is the Average Treatment Effect on the Treated (ATT, E[y_i1 – y_i0 | treat_i = 1]), because in our case, we didn't have a formal control group but instead, multiple groups treated simultaneously. Essentially, we estimated a heterogeneous treatment effect from an event study across different treated groups using a Diff-In-Diff-like setting. Note that because the treatment in our study was rather exogenous (as opposed to most other event studies, where the treatment was endogenous), the ATT in our study was actually equivalent to the Average Treatment Effect (ATE, E[y_i1 – y_i0]), because the chance of getting treatment was orthogonal to the outcome (treat_i ⊥ (yi1, yi0)) (e.g., Rosenbaum and Rubin (1983, 1984).
[39] Besides, we also conducted a survival analysis using the Cox-Proportional Hazard Model to examine the COVID-19 mean treatment effect on an individual's Hazard rate of opt out. Overall, we found that the mean effect remained qualitatively consistent.

To further understand the effect of political affiliation, we also ran a pooled analysis with ten blue cities and ten red cities. For robustness checking, we also conducted analyses using the Negative Binomial model with block-level and day-level fixed effects.

We provide the corresponding results in Tables 3-4. Next, we will discuss our findings in detail from these two levels of analysis.

## 6. Findings

In this section, we first discuss our main findings. Then, we will also discuss some additional falsification tests conducted to validate our findings.

### 6.1 Main Results

We estimated both individual-day-level and block-day-level models specified in Equations [1]-[3], and we report our empirical results in Tables 2-4.

Table 2 presents the results of exploring for the effects of the COVID-19 national emergency declaration, income, gender, and race on privacy concerns in D.C. on the individual-day level. We conducted the same econometric analysis on all of the cities listed above. The results demonstrated strong robustness across the different cities (i.e., they were qualitatively similar). We choose D.C. as an example here. The estimation result tables for the other cities are provided in Appendix A, Tables A1-A20.

**Table 2. Effects of COVID-19, Income, Gender, and Race on Privacy Concern**

| City = D.C.<br>Tr. = National Emergency | (1)<br>Treatment Effect<br>(Logit) | (2)<br>Interaction w/ Income<br>(Logit) | (3)<br>Interaction w/ Race<br>(Logit) | (4)<br>Interaction w/ Income<br>(LPM) | (5)<br>Interaction w/ Race<br>(LPM) |
|---|---|---|---|---|---|
| Treat | -0.027**<br>(0.011) | | | | |
| Treat × Income < 60K | | -0.382****<br>(0.035) | | -0.068***<br>(0.004) | |
| Treat × Income 60-100K | | -0.159****<br>(0.048) | | -0.016***<br>(0.004) | |
| Treat × Income 100-150K | | 0.131****<br>(0.041) | | 0.058***<br>(0.005) | |
| Treat × Income 150-200K | | 0.223****<br>(0.059) | | 0.015**<br>(0.006) | |
| Treat × Income > 200K | | 0.079**<br>(0.032) | | 0.060***<br>(0.003) | |
| Treat × Race White | | | 0.480****<br>(0.048) | | 0.035***<br>(0.006) |
| Treat × Race Black | | | 0.218****<br>(0.049) | | 0.011*<br>(0.006) |
| Treat × Race Asian | | | 1.516****<br>(0.091) | | 0.030***<br>(0.009) |
| Treat × Race Native | | | -0.059<br>(0.321) | | -0.018<br>(0.037) |
| Treat × Race Others | | | -0.443****<br>(0.046) | | 0.023***<br>(0.006) |
| Time Trend | 0.047***<br>(0.000) | 0.047***<br>(0.000) | 0.047***<br>(0.000) | 0.048***<br>(0.000) | 0.048***<br>(0.000) |
| Mobile App Usage | Yes | Yes | Yes | Yes | Yes |
| Week Fixed Effect | Yes | Yes | Yes | Yes | Yes |
| Day of Week Fixed Effect | Yes | Yes | Yes | Yes | Yes |
| Individual Fixed Effect | Yes | Yes | Yes | Yes | Yes |
| Log likelihood | -723950.51 | -723655.16 | -723477.58 | ---- | ---- |
| Observations | 1,542,977 | 1,542,977 | 1,542,977 | 1,542,977 | 1,542,977 |

\* $p < 0.10$, \*\* $p < 0.05$, \*\*\* $p < 0.01$, \*\*\*\* $p < 0.001$.

Column (1) shows the mean effect of the COVID-19 national emergency declaration on people's likelihood of opting out of location tracking. The negative and statistically significant coefficient with magnitude -0.027 suggests that the COVID-19 national emergency has a negative marginal effect and that on an average it will decrease the probability of a consumer opting out of location tracking by 1.6%. This result is consistent with the model-free evidence, and indicates that Americans are more willing to share their location data after the Trump Administration declared the national emergency.

Column (2) looks at how the treatment effect is heterogeneous across different income groups. The coefficients on the low-income bracket (below 60k) and 60k-100k are both negative and statistically significant, whereas the coefficients on the high-income brackets (100k-150k, 150k-200k, >200k) are positive and statistically significant. These results are striking. In general, they suggest that compared with low-income people, high-income people were not only more privacy-concerned before COVID-19, but also have become even more privacy-concerned and more likely to opt out of location tracking. In contrast, low-income people have become even less privacy-conscious after COVID-19. Column (3) reports the treatment effect as interacted with race. The findings regarding how race moderates the main effect are heterogeneous across different cities. Columns (4) and (5) present the analysis results based on the Linear Probability Model (LPM).

We also looked at how the treatment effect varies with gender. The result indicated that males, in general, are more privacy-concerned than females, and that moreover, after COVID-19, males have become even more privacy-concerned and more likely to opt out of location tracking. In contrast, females have become even less privacy-concerned. The above findings remain highly similar among the cities. The details across all 20 cities are reported in Appendix A, Tables A1-A20.

As a response to COVID-19, people have been encouraged to stay home and practice social distancing to help stem the spread of the viral pandemic. It was reported by CNN that the Trump administration wanted to use personal mobile location data to track the level of social distancing and trace close contacts to prevent the spread of coronavirus. An intriguing question given this situation is whether people may want to sacrifice their privacy for the greater social good. So, we turned to an examination of how people's opt-out decision is affected by the COVID-19 national emergency declaration and social distancing across the 20 cities. We present our results on Washington D.C. in Table 3 as an example. Results for all of the other cities are presented in Tables B1-B20 in Appendix B.

The analysis was conducted on the block level. We report in Column (1) of Table 3 a specification which includes the national emergency treatment (i.e., Treat), and three variables that capture how well people practice social distancing (i.e., Total Daily Contacts, Daily Travel Distance, and Daily Average Travel Speed). We were also interested in the heterogeneous treatment effect on people practicing different levels of social distancing. So, we included an additional specification with an interaction effect between Treat and Total Daily Contacts.

The coefficient on Treat, across the 20 cities, tells a consistent story: people were less likely to opt out and more willing to share their mobile locations after the COVID-19 national emergency declaration. The coefficients on Total Daily Contacts, Daily Travel Distance, and Daily Average Travel Speed are both positive and significant. This indicates a positive relationship between the practice of social distancing and the willingness to share location data. People who travelled more and interacted with a larger number of close contacts during the pandemic were also more likely to opt out of location tracking. On the flip side, people who practiced social distancing (i.e., those who travelled less and interacted with fewer close contacts during the pandemic) were also more likely to share their location data. Taking the coefficient on Total Daily Contacts in Table 3 as an example, we can explain the result in an intuitive way: one unit increase in face-to-face contact in a block, is associated with a 1.303 (=exp(0.265)) increase in the number of opt outs in the same block on the same day.

**Table 3. Effects of COVID-19 and Social Distancing on Privacy Concern**

| City = D.C.<br><br>Tr. = National Emergency | (1)<br>Mean Effect<br>(Poisson) | (2)<br>Interaction w/<br>Daily Contacts<br>(Poisson) | (3)<br>Mean Effect<br>(Negative Binomial) | (4)<br>Interaction w/<br>Daily Contacts<br>(Negative Binomial) |
|---|---|---|---|---|
| Treat | -0.262****<br>(0.028) | -0.416****<br>(0.029) | -0.239***<br>(0.037) | -0.350***<br>(0.039) |
| Treat × Total Daily Contacts |  | 0.219****<br>(0.012) |  | 0.193***<br>(0.020) |
| Total Daily Contacts | 0.265****<br>(0.006) | 0.206****<br>(0.007) | 0.272***<br>(0.010) | 0.218***<br>(0.011) |
| Daily Travel Distance | 0.358****<br>(0.010) | 0.363****<br>(0.010) | 0.410***<br>(0.014) | 0.403***<br>(0.014) |
| Daily Avg. Travel Speed | 0.081****<br>(0.001) | 0.081****<br>(0.001) | 0.102***<br>(0.002) | 0.101***<br>(0.002) |
| Time Trend | 0.016***<br>(0.000) | 0.017***<br>(0.000) | 0.020***<br>(0.001) | 0.021***<br>(0.001) |
| Control Variables |  |  |  |  |
| Population | Yes | Yes | Yes | Yes |
| Block Land Area | Yes | Yes | Yes | Yes |
| Income | Yes | Yes | Yes | Yes |
| Gender | Yes | Yes | Yes | Yes |
| Race | Yes | Yes | Yes | Yes |
| Number of Existing Users | Yes | Yes | Yes | Yes |
| Number of Opt-in Users | Yes | Yes | Yes | Yes |
| Mobile App Usage | Yes | Yes | Yes | Yes |
| Week Fixed Effect | Yes | Yes | Yes | Yes |
| Day of Week Fixed Effect | Yes | Yes | Yes | Yes |
| Log likelihood | -44233.39 | -43823.91 | -40542.02 | -40493.94 |
| Observations | 84,270 | 84,270 | 84,270 | 84,270 |

\* $p < 0.10$, \*\* $p < 0.05$, \*\*\* $p < 0.01$, \*\*\*\* $p < 0.001$.
Models (1) ~ (2) are based on the Poisson Model; Models (3) ~ (4) are based on the Negative Binomial Model.

**Table 4. Pooled Analyses: Red Cities vs. Blue Cities**
**Effects of COVID-19 National Emergency Declaration and Social Distancing on Privacy Concern**

| Tr. = National Emergency | (1) Main Effect (10 Red Cities) | (2) w/ Social Distancing (10 Red Cities) | (3) Interaction w/ Daily Contacts (10 Red Cities) | (4) Main Effect (10 Blue Cities) | (5) w/ Social Distancing (10 Blue Cities) | (6) Interaction w/ Daily Contacts (10 Blue Cities) | (7) Main Effect Using Pooled Regression w/ Blue Dummy (All 20 Cities) |
|---|---|---|---|---|---|---|---|
| Treat | -0.127**** (0.010) | -0.112**** (0.010) | -0.511**** (0.012) | -0.325**** (0.010) | -0.264**** (0.011) | -0.472**** (0.011) | -0.306**** (0.007) |
| Treat × Daily Contacts | | | 0.222**** (0.003) | | | 0.315**** (0.005) | |
| Daily Contacts | | 0.431**** (0.002) | 0.326**** (0.002) | | 0.311**** (0.003) | 0.221**** (0.003) | |
| Daily Travel Distance | | 0.876**** (0.004) | 0.859**** (0.004) | | 0.975**** (0.004) | 0.970**** (0.004) | |
| Blue City Dummy | | | | | | | -0.399**** (0.004) |
| Time Trend | 0.008**** (0.000) | 0.025**** (0.000) | 0.024**** (0.000) | 0.015**** (0.000) | 0.023**** (0.000) | 0.024**** (0.000) | 0.018**** (0.000) |
| Inflection Rate | -4.858**** (0.188) | -4.015**** (0.185) | -2.097**** (0.185) | -1.913**** (0.046) | -0.975**** (0.046) | -0.626**** (0.047) | -2.059**** (0.041) |
| Death Rate | -13.53**** (0.427) | -12.01**** (0.397) | -9.645**** (0.382) | -4.157**** (0.110) | -4.946**** (0.111) | -4.959**** (0.111) | -3.381*** (0.092) |
| *Control Variables* | | | | | | | |
| Population, Block Land Area | Yes | Yes | Yes | Yes | Yes | Yes | Yes |
| Income, Gender, Race | Yes | Yes | Yes | Yes | Yes | Yes | Yes |
| Number of Existing Users | Yes | Yes | Yes | Yes | Yes | Yes | Yes |
| Number of Opt-in Users | Yes | Yes | Yes | Yes | Yes | Yes | Yes |
| Mobile App Usage | Yes | Yes | Yes | Yes | Yes | Yes | Yes |
| Week, Day of Week Fixed Effect | Yes | Yes | Yes | Yes | Yes | Yes | Yes |
| Log likelihood | -354744.31 | -309058.27 | -306768.31 | -409090.70 | -371947.60 | -370188.06 | -775273.14 |
| Observations | 601,974 | 601,974 | 601,974 | 1,165,364 | 1,165,364 | 1,165,364 | 1,767,338 |

\* $p < 0.10$, ** $p < 0.05$, *** $p < 0.01$, **** $p < 0.001$.
Models (1)~(7) are based on the Poisson Model. We also ran analyses using the Negative Binomial Model, and the results remained highly consistent.

What is more striking was that the positive relationship between social distancing and location sharing became even more salient after the national emergency was declared. This is supported by the positive and statistically significant coefficient (i.e., 0.219) on the interaction terms Treat and Total Daily Contacts in Column (2) of Table 3.

Because the outcome variable was the number of total opt outs in each block, we conducted count data analyses using both the Poisson and Negative Binomial models. Columns (1) – (2) are based on the Poisson Model and Columns (3) – (4) are based on the Negative Binomial Model. We found that the results remained highly consistent.[40] While Table 3 shows the results for D.C., Appendix B provides the replicated-analysis results for all 20 cities. The results were robust across cities as well.

**Table 5 – Effects of National Emergency Declaration vs. State-specific Lockdown Order on Privacy Concern**

| City | State | State Order Effective Date | Mean Effect (State-specific) | Mean Effect (National Emergency) |
|---|---|---|---|---|
| SFO | CA | Mar 19, 2020 | -0.708**** (0.029) | -0.678**** (0.030) |
| New York City | NY | Mar 22, 2020 | -0.828** (0.028) | -0.344**** (0.027) |
| New Orleans | LA | Mar 23, 2020 | -0.489**** (0.032) | -0.048* (0.028) |
| Seattle | WA | Mar 23, 2020 | 0.106**** (0.031) | -0.469**** (0.031) |
| Boston | MA | Mar 24, 2020 | -0.468**** (0.025) | -0.562**** (0.026) |
| Lexington | KY | Mar 26, 2020 | -0.095**** (0.024) | -0.580**** (0.026) |
| Colorado Springs | CO | Mar 26, 2020 | -0.015 (0.025) | -0.678**** (0.026) |
| Oklahoma City | OK | Mar 28, 2020 | -0.059* (0.030) | -0.188**** (0.027) |
| Virginia Beach | VA | Mar 30, 2020 | 0.189**** (0.032) | -0.543**** (0.029) |
| Baltimore | MD | Mar 30, 2020 | -0.271**** (0.026) | -0.337**** (0.027) |
| Wichita | KS | Mar 30, 2020 | 0.050 (0.031) | -0.239**** (0.027) |
| Nashville | TN | Mar 31, 2020 | -0.147**** (0.030) | -0.264**** (0.027) |
| Phoenix | AZ | Mar 31, 2020 | -0.060* (0.036) | -0.441**** (0.030) |
| D.C. | DC | Apr 1, 2020 | -0.075*** (0.027) | -0.279**** (0.014) |
| Pittsburgh | PA | Apr 1, 2020 | -0.001 (0.033) | -0.457**** (0.029) |
| Philadelphia | PA | Apr 1, 2020 | -0.122**** (0.030) | -0.476**** (0.029) |
| Austin | TX | Apr 2, 2020 | 0.032 (0.033) | -0.312**** (0.027) |
| Arlington | TX | Apr 2, 2020 | 0.076*** (0.028) | -0.215**** (0.026) |
| Jacksonville | FL | Apr 3, 2020 | 0.044 (0.035) | -0.254**** (0.030) |
| Omaha | NE | n/a# | -- | -0.392**** (0.028) |
| Same control variable used in Table 3 | | | Yes | |

\* $p < 0.10$, \*\* $p < 0.05$, \*\*\* $p < 0.01$, \*\*\*\* $p < 0.001$.
# Nebraska never ordered residents to stay home.
The results are based on the main analyses using the Poisson Model. We also ran the same analyses using the Negative Binomial Model, and the results remained highly consistent.

---

[40] As a robustness test, we also ran the same analysis by controlling for the social demographics, population density, app usage, and opt-in/out activities of the neighboring (i.e., geographically adjacent) blocks. We found that our results remained highly consistent. The detailed results are provided in Table C in Appendix C.

Furthermore, we conduct pooled analyses for all 20 cities to see if the effect of social distancing on privacy concerns varied across political affiliations. The results are shown in Table 4. First, we ran the pooled analysis for all 10 red cities and for all 10 blue cities separately, as shown in Columns (1) to (6). Then, we ran the pooled analysis with all 20 cities together using a "Blue City" indicator, as shown in Column (7). Looking at Columns (1) to (6), the coefficients are qualitatively similar to those in Table 3. This indicates that individuals who practice social distancing are more likely to share their location data than are those who do not practice social distancing. Besides, comparing the coefficients of the interaction effect between treatment and daily contacts in Columns (3) and (6) of Table 4, we found that such a positive correlation between social distancing and location sharing appears more salient in blue cities than in red ones after COVID-19.

In addition, we conducted similar analyses at the block-day level for all 20 cities using the state-specific lockdowns in each city as the treatment variable. We provide the mean effects of the two treatments (National Emergency vs. State Lockdown) in Table 5 for comparison. We found that the results remained qualitatively consistent in a large majority of the cities.[41]

## 6.2 Summary of Findings

In consideration of all the empirical evidence shown above, we summarize our major findings as follows.

**Political Affiliation and Privacy Concern**. People were more willing to share their mobile locations after President Trump declared the COVID-19 national emergency on March 13. The results were also consistent in each city when we used the state-specific lockdowns in each city as the treatment variable. Despite the increasing concern that government authorities, the private sector, and public health experts may have to use individual-level location data to track the coronavirus, we find that there is a significant and decreasing trend of opt out of location sharing with mobile apps in the U.S. Our results are consistent with recent studies[42] demonstrating an increased willingness of Americans to share location and health data to help slow the spread of the virus and reduce the lockdown period. While in general people in the blue cities were more privacy-concerned than those in the red cities before the advent of the COVID-19 crisis, there was a significant decrease in opt-out rates after COVID-19, and this effect was more salient in the blue cities than in the red cities.

---

[41] We noticed the effect of state lockdown was positive in Arlington and Virginia Beach. This was likely due to the fact that the lockdown orders in those two states (TX and VA) were issued relatively late (Apr. 2 and Mar 30); we did not have enough daily observations to fully observe the post-treatment trend (i.e., April 15 is the last day of observation in the current data set). The estimate of the effect of the state order also appears positive for Seattle, which seems somewhat noisy. We are now collecting post-April 15 data to run additional robustness tests.
[42] https://blogs.scientificamerican.com/observations/will-americans-be-willing-to-install-covid-19-tracking-apps/
http://webuse.org/covid/
https://www.emarketer.com/content/consumers-are-more-willing-to-share-private-data-during-covid-19

**Social Distancing and Privacy Concern.** The practice of social distancing and willingness to share location data were positively correlated: people who practice social distancing (i.e., those who travel less and interact with fewer close contacts during the pandemic) were also more likely to share their location data, whereas anti-social-distancing people (i.e., those who travel more and interact with a greater number of close contacts during the pandemic) were more likely to opt out of location tracking; such positive relationship between social distancing and location sharing has become even more salient after Trump declared the national emergency (and after the state-specific lockdowns); such positive relationship between social distancing and location sharing appeared to be more salient in blue cities than in red ones after the declaration.

**Social Demographics and Privacy Concern.** High-income people and males, in general, are more privacy-concerned than are low-income people and females. After COVID-19, high-income people has become even more privacy-concerned and more likely to opt out of location tracking.

Finally, an interesting observation from both our individual-level analyses (Table 2, Tables A1-A20) and block-level analyses (Table 3, Tables B1-B20) suggests two opposing forces with respect to privacy concern. On the one hand, our findings show a positive relationship between willingness to practice social distancing and willingness to share location data. This seems also to suggest an underlying pro-social cause that might positively affect both people's compliance to national/state lockdown orders and their willingness to trade privacy for social good during the pandemic period. Such pro-social behavior is also moderated by the potential health risk. Previous research[43] has shown that people whom the CDC has identified as being higher risk were more likely to be willing to install contact-tracing apps for location tracking. We found consistent evidence that health risk factors (infection rate and death rate) had a significant negative effect on people's opt-out behavior, and that cities that were hit harder by COVID-19 demonstrated a more salient drop in the opt-out rate[44].

On the other hand, the change in people's opt-out behavior could also be a result of a change in time availability. With the increased time available due to shelter-at-home and social-distancing policies, it is conceivable that consumers may more likely have the time to read through the mobile apps' privacy policies and react accordingly given the increased surveillance concerns and awareness. In our data, we observe a general increasing trend in people opting out of location tracking before COVID-19. If indeed people had more time at home after COVID-19 and were checking their privacy settings more thoroughly, we would expect to see more opt outs, given the general increasing trend in privacy concern. Hence, this indicates that time availability may negatively affect people's willingness to share location data. We also saw such evidence from the weekend effect in the data – when people have more free time during weekends they are more likely to opt out of location tracking compared with during weekdays. This pattern becomes less prominent after COVID-19, potentially due to a blurred boundary between weekends and weekdays, particularly after the stay-at-home order. Therefore, whether a consumer is more likely to share location data or not after COVID-19 depends on which force plays the dominant role.

---

[43] http://webuse.org/covid/
[44] We found that blue cities on average were hit harder by COVID-19 than red cities in terms of both infection rate and death rate (as shown in Table D in Appendix D). This is consistent with our observation of a pro-social cause and to some extent explains why blue cities witnessed a larger drop in opt-out rate after COVID.

Interestingly, our results show that on average people were more likely to share their location data after COVID-19. This indicates that the pro-social cause seems to play a more dominant role in the entire American population. This trend is especially prominent for the low-income groups. This could potentially be due to the fact that many low-income people are essential workers and did not experience a marked increase in time availability after COVID-19. In contrast, we observed that high-income people became more privacy-concerned and more likely to opt out of location tracking. This seems to indicate that the effect of time availability plays a more dominant role in this case— high-income people having received a stronger positive shock to their time availability after COVID-19. Given the more practical nature of high-income groups, they would be more likely to use such increased time availability to check/revise privacy settings.[45]

## 6.3 Falsification Tests

To validate the robustness of our findings, we conducted a number of falsification tests.

**(1) Alternative Treatment Dates.**

First, we were interested in testing whether the treatment (the COVID-19 national emergency declaration on Mar. 13) indeed has brought a negative shock to people's privacy opt-out behavior, or if there was simply a downward trend over time. The week, day-of-week dummies and day index in our model could help to a large extent to control for such time trends. Besides, as we can see in Figure 3, the predominant drop in the opt-out rate does not appear until after the treatment.

To further validate our results, we reran our analyses using alternative hypothetical dates prior to March 13 as treatment dates. Because March 13 is a Friday, we chose two alternative Fridays in January (Jan. 17, Jan. 24) and two in February (Feb. 14, Feb. 28), and ran our estimations, separately, again. Overall, we did not find a similar statistically significant drop in people's opt-out behavior under the hypothetical treatment dates.

---

[45] Our finding is also consistent with several recent studies (e.g., Chiou and Tucker 2020) showing that people from higher-income regions display more compliance with social distancing and "remain-at-home" orders (hence enjoying a higher increase in time availability).

We conducted this falsification test for all of the cities. We illustrate the results in Table 6 using D.C. as an example. We can see that the estimated treatment effects are mostly statistically insignificant, and moreover, the direction of the effects is mostly positive (as opposed to a negative effect after the actual treatment of Mar. 13).

**(2) Exogeneity of Treatment.**

Second, there might be a concern that the treatment is not completely exogenous. For example, the mainstream media articles on location tracking could have appeared because they had observed significant changes in people's willingness to share their location data. We argue that this is very unlikely. First of all, those articles were written in response to the recent emergence of contact-tracing technologies developed by tech companies like Google and Apple, instead of in reaction to the observation of changes in consumers' opt-out behavior. Furthermore, the development of contact-tracing technologies has been in response to the COVID-19 pandemic, which is completely exogenous and has nothing to do with consumer privacy behavior. In addition, to further test this, we also ran a set of prediction models to see if people's opt-out behavior can predict the treatment variable. We found the predictive performance to be rather low (e.g., Taking Washington D.C. as an example, the correlation between previous opt-out rate and future treatment was statistically insignificant. The prediction accuracy was merely 67%).

**Table 6. Falsification Test - Alternative Treatment Dates**

| City = D.C.<br>Tr. = National Emergency | (1)<br>Treatment<br>Jan. 17 | (2)<br>Treatment<br>Jan. 24 | (3)<br>Treatment<br>Feb. 14 | (4)<br>Treatment<br>Feb. 28 |
|---|---|---|---|---|
| Treat | 0.069* | 0.051 | -0.006 | 0.038 |
|  | (0.040) | (0.033) | (0.026) | (0.024) |
| Control Variables | | | | |
| Population | Yes | Yes | Yes | Yes |
| Block Land Area | Yes | Yes | Yes | Yes |
| Population Income | Yes | Yes | Yes | Yes |
| Population Gender | Yes | Yes | Yes | Yes |
| Population Race | Yes | Yes | Yes | Yes |
| Number of Existing Users | Yes | Yes | Yes | Yes |
| Number of Opt-in Users | Yes | Yes | Yes | Yes |
| Mobile App Usage | Yes | Yes | Yes | Yes |
| Week Fixed Effect | Yes | Yes | Yes | Yes |
| Day of Week Fixed Effect | Yes | Yes | Yes | Yes |
| Log likelihood | -48213.46 | -48213.73 | -48214.89 | -48213.72 |
| Observations | 84,270 | 84,270 | 84,270 | 84,270 |

* $p < 0.10$, ** $p < 0.05$, *** $p < 0.01$, **** $p < 0.001$.

**(3) Opt outs vs. Uninstallation/New Installation**

Finally, a potential alternative explanation to our main finding is that the observed decreasing trend of opt outs in the post-treatment period could have been due to an overall decrease in people's mobile app usage. In particular, people could simply uninstall an app due to lack of interest in it, instead of just opting out of location tracking. The uninstallation behavior could also lead to the cessation of location data. However, it is important to note that in order to completely disappear from our location-data set, the individual needs to simultaneously uninstall all of the affiliated apps that use the location tracking SDK provided by our location-data partner. Given the market share of the company and the coverage of their client apps, this is very unlikely to occur.

A related confounder is that the observed drop in opt outs after COVID-19 could also have been due to a sudden increase of new app installations— that is, because more people were stuck at home, they could have been installing new apps for convenience, some of which may require location tracking to function (e.g., Uber Eats for food delivery). This could potentially lead to more opt ins to location sharing, but it would have nothing to do with people's existing privacy perspectives.

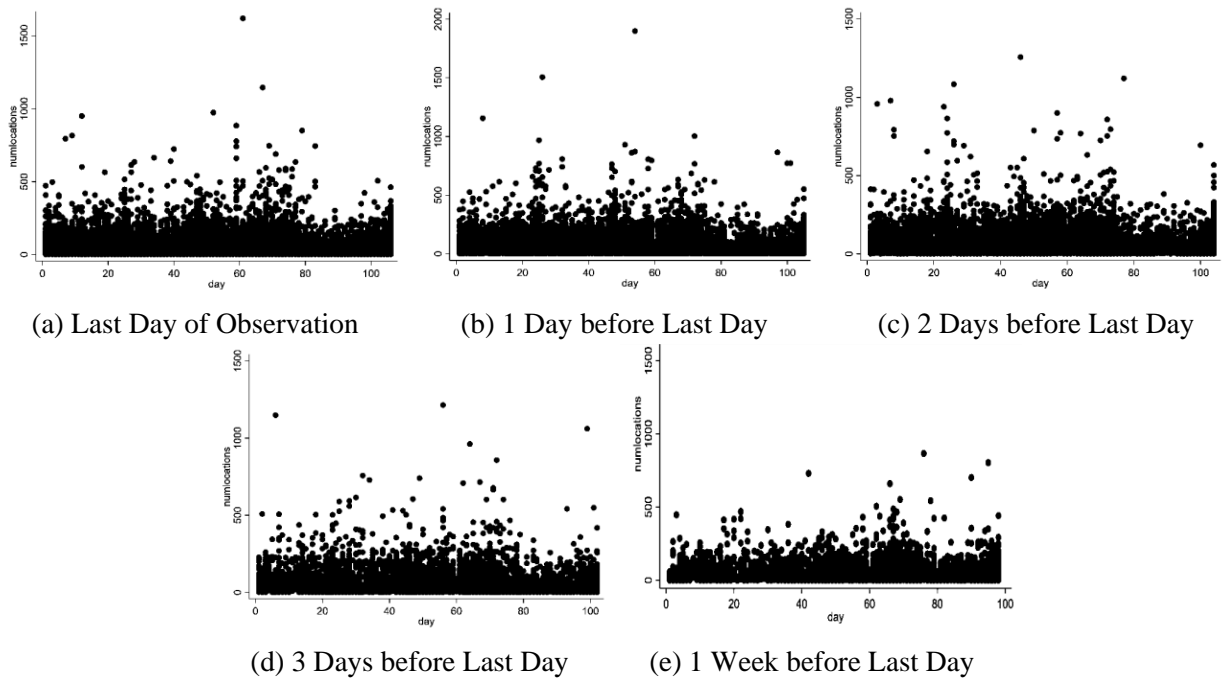

(a) Last Day of Observation　　(b) 1 Day before Last Day　　(c) 2 Days before Last Day

(d) 3 Days before Last Day　　(e) 1 Week before Last Day

**Figure 8. Distribution of total #app activities for each "opt-out" user on (a) last day of observing that user; (b) 1 day before last day; (c) 2 days before last day; (d) 3 days before last day; and (e) 1 week before last day**

To further test these alternative explanations, we conducted robustness tests to check individuals' app usage behavior before the last day of observation (i.e., before the user opted out). Specifically, if our result were due to uninstallations of all of the affiliated apps, we would expect to see a pre-existing decreasing trend for the opt-out users in their daily app activities before the opt out. In contrast, if our result were due to an increase in the new app installation, we would expect to see a pre-existing increasing trend in the users' daily app activities. In our main analyses, we explicitly controlled for all of the app-usage-related factors (e.g., total usage time, unique app categories, daily number of locations) over time at both the individual and block levels, and we found our findings to be highly robust. This step to a large extent controls for the pre-existing trends in app usage behavior. Moreover, based on a further comparison of the distribution of individual-level app activities, we did not find either a decreasing or an increasing trend before the opt out. For illustration, we show, in Figure 8, the distributions of the total individual-level numbers of app activities for Washing D.C.[46] For each opt-out user, we plot the total app activities on the last day of observation, one/two/three days before the last day, and one week before the last day. As we can see in Figure 8, the distribution remains highly consistent across all days, suggesting no significant pre-existing trend in user app activities.

## 7. Limitations

Our paper has some limitations. First, we cannot necessarily ascertain if the privacy opt-out behavior we observed is because of pro-social behavior or because people had more available time while they were at home that might make them want to change their privacy choices. That being said, we did observe that there was a general increasing trend of people opting out of location tracking over time before the COVID-19 crisis, but that after the national emergency declaration and the state-specific lockdown dates, there was a reduction in the opt-out rates. If indeed people had more time at home after COVID-19 and were reading the privacy policies of mobile apps and checking their privacy settings more thoroughly, we would have expected to see more opt outs, given the general increasing trend in privacy concern. But we in fact saw fewer opt outs, and so this partially mitigates the concern about time, as opposed to pro-social behavior, being the driving factor. However, we do acknowledge that to unpack the mechanisms, we will likely need to conduct a survey that is outside the scope of this paper. We leave it to future research.

## 8. Conclusions

Having used a granular individual-level dataset consisting of over 22 billion records from ten 'Blue' (Democratic) and ten 'Red' (Republican) U.S. cities, we present some of the first evidence of how Americans responded to increasing privacy concerns during the COVID-19 pandemic. We demonstrate a significant decrease in the opt out of location sharing in the U.S. While areas with more Democrats were more privacy-concerned than areas with more Republicans before the advent of the COVID-19 crisis, there was a decrease in the overall opt-out rates after COVID-19, and this effect was more salient among the Democratic than Republican cities. People who practice social distancing were also less likely to opt out, whereas the converse was true for those less compliant with social distancing. This effect appeared to be more salient among the Democratic than Republican cities. Finally, we show that high-income populations, cities with a higher proportion of the white population, and males in general were more privacy-concerned relative to low-income populations, cities with a more diverse population, and females. Thus, demographic differences influenced the extent of pro-social behavior exhibited by Americans.

---

[46] We have also provided a similar analysis of total app usage time in Figure E in Appendix E.

Our results are consistent with recent surveys held in the post COVID-19 period that indicate that a majority of American are willing to allow mobile apps to disclose their locations in order to help public officials flag hot spots of COVID-19 and help slow the spread of the virus thereby. Our work can help dictate where more physical resources may be necessary for local governments to invest in order to curb the spread of Covid-19 in places that people are more likely to opt-out from location data sharing. Overall, our research demonstrates that Americans in both Blue and Red cities generally formed a unified front in sacrificing personal privacy for the societal good amid COVID-19, while simultaneously exhibiting a divergence in the extent of such a sacrifice along the lines of political affiliation, social-distancing compliance, and demographics.

# Appendix A – Per City Results [1]
# Effect of COVID-19 and Demographics on Privacy Concern - Individual Level Analyses

### Table A1 - Boston

| City = Boston<br>Tr. = National Emergency | (1)<br>Main Effect<br>(Logit) | (2)<br>Interaction<br>w/ Income<br>(Logit) | (3)<br>Interaction<br>w/ Gender<br>(Logit) | (4)<br>Interaction<br>w/ Race<br>(Logit) | (5)<br>Interaction<br>w/ Income<br>(LPM) | (6)<br>Interaction<br>w/ Gender<br>(LPM) | (7)<br>Interaction<br>w/ Race<br>(LPM) |
|---|---|---|---|---|---|---|---|
| Treat | -0.004<br>(0.011) | | | | | | |
| Treat ×<br>Income less than 60K | | -0.211****<br>(0.043) | | | -0.052****<br>(0.003) | | |
| Treat ×<br>Income 60K – 100K | | -0.354****<br>(0.063) | | | -0.109****<br>(0.006) | | |
| Treat ×<br>Income 100K – 150K | | -0.782****<br>(0.077) | | | -0.092****<br>(0.007) | | |
| Treat ×<br>Income 150K – 200K | | 0.581****<br>(0.087) | | | 0.268****<br>(0.007) | | |
| Treat ×<br>Income more than 200K | | 0.216****<br>(0.041) | | | 0.082****<br>(0.002) | | |
| Treat ×<br>Female | | | -0.010<br>(0.029) | | | 0.034****<br>(0.002) | |
| Treat ×<br>Male | | | 0.013<br>(0.056) | | | 0.038****<br>(0.005) | |
| Treat ×<br>Race Others | | | | 0.091**<br>(0.042) | | | 0.079****<br>(0.003) |
| Treat ×<br>Race White | | | | -0.053<br>(0.045) | | | 0.001<br>(0.003) |
| Treat ×<br>Race Black | | | | -0.718****<br>(0.054) | | | -0.192****<br>(0.004) |
| Treat ×<br>Race Asian | | | | 0.020<br>(0.056) | | | -0.065****<br>(0.005) |
| Treat ×<br>Race Native | | | | 3.163****<br>(0.446) | | | -0.251****<br>(0.040) |
| Mobile App Usage Controlled | Yes | Yes | Yes | Yes | Yes | Yes | Yes |
| Week Fixed Effect | Yes | Yes | Yes | Yes | Yes | Yes | Yes |
| Day of Week Fixed Effect | Yes | Yes | Yes | Yes | Yes | Yes | Yes |
| Individual Fixed Effect | Yes | Yes | Yes | Yes | Yes | Yes | Yes |
| Log likelihood /<br>R-squared | -727156.46 | -727021.97 | -727156.44 | -726842.64 | 0.461 | 0.460 | 0.462 |
| Observations | 1,501,913 | 1,501,913 | 1,501,913 | 1,501,913 | 1,501,913 | 1,501,913 | 1,501,913 |

\* p < 0.10, \*\* p < 0.05, \*\*\* p < 0.01, \*\*\*\* p < 0.001.

**Table A2 – D.C.**

| City = D.C.<br>Tr. = National Emergency | (1)<br>Main Effect<br>(Logit) | (2)<br>Interaction<br>w/ Income<br>(Logit) | (3)<br>Interaction<br>w/ Gender<br>(Logit) | (4)<br>Interaction<br>w/ Race<br>(Logit) | (5)<br>Interaction<br>w/ Income<br>(LPM) | (6)<br>Interaction<br>w/ Gender<br>(LPM) | (7)<br>Interaction<br>w/ Race<br>(LPM) |
|---|---|---|---|---|---|---|---|
| Treat | -0.027**<br>(0.011) | | | | | | |
| Treat ×<br>Income less than 60K | | -0.382****<br>(0.035) | | | -0.068***<br>(0.004) | | |
| Treat ×<br>Income 60K – 100K | | -0.159****<br>(0.048) | | | -0.016***<br>(0.004) | | |
| Treat ×<br>Income 100K – 150K | | 0.131****<br>(0.041) | | | 0.058***<br>(0.005) | | |
| Treat ×<br>Income 150K – 200K | | 0.223****<br>(0.059 | | | 0.015**<br>(0.006) | | |
| Treat ×<br>Income more than 200K | | 0.079**<br>(0.032) | | | 0.060***<br>(0.003) | | |
| Treat ×<br>Female | | | -0.234****<br>(0.032) | | | -0.057*<br>(0.003) | |
| Treat ×<br>Male | | | 0.433****<br>(0.062) | | | -0.013****<br>(0.007) | |
| Treat ×<br>Race Others | | | | -0.443****<br>(0.046) | | | 0.035***<br>(0.006) |
| Treat ×<br>Race White | | | | 0.480****<br>(0.048) | | | 0.011*<br>(0.006) |
| Treat ×<br>Race Black | | | | 0.218****<br>(0.049) | | | 0.030***<br>(0.009) |
| Treat ×<br>Race Asian | | | | 1.516****<br>(0.091) | | | -0.018<br>(0.037) |
| Treat ×<br>Race Native | | | | -0.059<br>(0.321) | | | 0.023***<br>(0.006) |
| Mobile App Usage Controlled | Yes | Yes | Yes | Yes | Yes | Yes | Yes |
| Week Fixed Effect | Yes | Yes | Yes | Yes | Yes | Yes | Yes |
| Day of Week Fixed Effect | Yes | Yes | Yes | Yes | Yes | Yes | Yes |
| Individual Fixed Effect | Yes | Yes | Yes | Yes | Yes | Yes | Yes |
| Log likelihood /<br>R-squared | -723950.51 | -723655.16 | -723927.00 | -723477.58 | 0.466 | 0.465 | 0.466 |
| Observations | 1,542,977 | 1,542,977 | 1,542,977 | 1,542,977 | 1,542,977 | 1,542,977 | 1,542,977 |

* $p < 0.10$, ** $p < 0.05$, *** $p < 0.01$, **** $p < 0.001$.

**Table A3 – Baltimore**

| City = Baltimore<br>Tr. = National Emergency | (1)<br>Main Effect<br>(Logit) | (2)<br>Interaction<br>w/ Income<br>(Logit) | (3)<br>Interaction<br>w/ Gender<br>(Logit) | (4)<br>Interaction<br>w/ Race<br>(Logit) | (5)<br>Interaction w/<br>Income<br>(LPM) | (6)<br>Interaction<br>w/ Gender<br>(LPM) | (7)<br>Interaction<br>w/ Race<br>(LPM) |
|---|---|---|---|---|---|---|---|
| Treat | 0.070****<br>(0.010) | | | | | | |
| Treat ×<br>Income less than 60K | | -0.516****<br>(0.084) | | | -0.062***<br>(0.003) | | |
| Treat ×<br>Income 60K – 100K | | -0.167*<br>(0.094) | | | 0.057****<br>(0.005) | | |
| Treat ×<br>Income 100K – 150K | | -0.138<br>(0.093) | | | 0.124****<br>(0.005) | | |
| Treat ×<br>Income 150K – 200K | | -0.020<br>(0.121) | | | 0.056****<br>(0.009) | | |
| Treat ×<br>Income more than 200K | | 0.387****<br>(0.084) | | | 0.050****<br>(0.003) | | |
| Treat ×<br>Female | | | -0.203****<br>(0.026) | | | 0.029****<br>(0.002) | |
| Treat ×<br>Male | | | 0.558****<br>(0.049) | | | 0.056****<br>(0.004) | |
| Treat ×<br>Race Others | | | | -0.516****<br>(0.063) | | | 0.039****<br>0.003) |
| Treat ×<br>Race White | | | | 0.706****<br>(0.065) | | | 0.058****<br>0.003) |
| Treat ×<br>Race Black | | | | 0.237****<br>(0.065) | | | -0.097****<br>0.003) |
| Treat ×<br>Race Asian | | | | 1.056****<br>(0.082) | | | 0.074****<br>(0.006) |
| Treat ×<br>Race Native | | | | 1.746****<br>(0.369) | | | -0.361****<br>(0.047) |
| Mobile App Usage Controlled | Yes | Yes | Yes | Yes | Yes | Yes | Yes |
| Week Fixed Effect | Yes | Yes | Yes | Yes | Yes | Yes | Yes |
| Day of Week Fixed Effect | Yes | Yes | Yes | Yes | Yes | Yes | Yes |
| Individual Fixed Effect | Yes | Yes | Yes | Yes | Yes | Yes | Yes |
| Log likelihood / R-squared | -777006.12 | -776778.3 | -776941.71 | -776417.23 | 0.418 | 0.417 | 0.419 |
| Observations | 1,540,515 | 1,540,515 | 1,540,515 | 1,540,515 | 1,540,515 | 1,540,515 | 1,540,515 |

\* $p < 0.10$, ** $p < 0.05$, *** $p < 0.01$, **** $p < 0.001$.

**Table A4 – Lexington**

| City = Lexington<br>Tr. = National Emergency | (1)<br>Main Effect<br>(Logit) | (2)<br>Interaction<br>w/ Income<br>(Logit) | (3)<br>Interaction<br>w/ Gender<br>(Logit) | (4)<br>Interaction<br>w/ Race<br>(Logit) | (5)<br>Interaction<br>w/ Income<br>(LPM) | (6)<br>Interaction<br>w/ Gender<br>(LPM) | (7)<br>Interaction<br>w/ Race<br>(LPM) |
|---|---|---|---|---|---|---|---|
| Treat | -0.176****<br>(0.009) | | | | | | |
| Treat × Income less than 60K | | -0.023<br>(0.073) | | | -0.142****<br>(0.006) | | |
| Treat × Income 60K – 100K | | -0.488****<br>(0.079) | | | -0.181****<br>(0.008) | | |
| Treat × Income 100K – 150K | | -0.593****<br>(0.106) | | | -0.208****<br>(0.010) | | |
| Treat × Income 150K – 200K | | 0.308**<br>(0.152) | | | -0.148****<br>(0.015) | | |
| Treat × Income more than 200K | | 0.016<br>(0.071) | | | 0.172****<br>(0.006) | | |
| Treat × Female | | | -0.333****<br>(0.037) | | | -0.046****<br>(0.004) | |
| Treat × Male | | | 0.314****<br>(0.071) | | | 0.135****<br>(0.008) | |
| Treat × Race Others | | | | -0.534****<br>(0.071) | | | 0.101****<br>(0.006) |
| Treat × Race White | | | | 0.326****<br>(0.074) | | | -0.085****<br>(0.006) |
| Treat × Race Black | | | | 0.534****<br>(0.088) | | | -0.088****<br>(0.008) |
| Treat × Race Asian | | | | 1.286****<br>(0.117) | | | -0.057****<br>(0.013) |
| Treat × Race Native | | | | 2.371****<br>(0.489) | | | -0.054****<br>(0.055) |
| Mobile App Usage Controlled | Yes | Yes | Yes | Yes | Yes | Yes | Yes |
| Week Fixed Effect | Yes | Yes | Yes | Yes | Yes | Yes | Yes |
| Day of Week Fixed Effect | Yes | Yes | Yes | Yes | Yes | Yes | Yes |
| Individual Fixed Effect | Yes | Yes | Yes | Yes | Yes | Yes | Yes |
| Log likelihood / R-squared | -738608.30 | -738452.41 | -738598.53 | -738526.83 | 0.437 | 0.437 | 0.438 |
| Observations | 1,462,140 | 1,462,140 | 1,462,140 | 1,462,140 | 1,462,140 | 1,462,140 | 1,462,140 |

\* $p < 0.10$, ** $p < 0.05$, *** $p < 0.01$, **** $p < 0.001$.

**Table A5 – Colorado Spring**

| City = Colorado Spring<br>Tr. = National Emergency | (1)<br>Main Effect<br>(Logit) | (2)<br>Interaction<br>w/ Income<br>(Logit) | (3)<br>Interaction<br>w/ Gender<br>(Logit) | (4)<br>Interaction<br>w/ Race<br>(Logit) | (5)<br>Interaction<br>w/ Income<br>(LPM) | (6)<br>Interaction<br>w/ Gender<br>(LPM) | (7)<br>Interaction<br>w/ Race<br>(LPM) |
|---|---|---|---|---|---|---|---|
| Treat | -0.078****<br>(0.009) | | | | | | |
| Treat ×<br>Income less than 60K | | -0.158<br>(0.114) | | | -0.008<br>(0.010) | | |
| Treat ×<br>Income 60K – 100K | | -0.336***<br>(0.117) | | | -0.053****<br>(0.010) | | |
| Treat ×<br>Income 100K – 150K | | -1.122****<br>(0.144) | | | -0.212****<br>(0.013) | | |
| Treat ×<br>Income 150K – 200K | | 0.293<br>(0.205) | | | 0.121****<br>(0.019) | | |
| Treat ×<br>Income more than 200K | | 0.211*<br>(0.113) | | | 0.055****<br>(0.009) | | |
| Treat ×<br>Female | | | -0.399****<br>(0.044) | | | -0.047****<br>(0.004) | |
| Treat ×<br>Male | | | 0.634****<br>(0.084) | | | 0.129****<br>(0.008) | |
| Treat ×<br>Race Others | | | | -0.396****<br>(0.058) | | | -0.103****<br>(0.005) |
| Treat ×<br>Race White | | | | 0.420****<br>(0.061) | | | 0.150****<br>(0.006) |
| Treat ×<br>Race Black | | | | 0.284***<br>(0.106) | | | 0.105****<br>(0.010) |
| Treat ×<br>Race Asian | | | | -2.086****<br>(0.189) | | | -0.289****<br>(0.018) |
| Treat ×<br>Race Native | | | | 2.491****<br>(0.462) | | | 0.344****<br>(0.042) |
| Mobile App Usage<br>Controlled | Yes | Yes | Yes | Yes | Yes | Yes | Yes |
| Week<br>Fixed Effect | Yes | Yes | Yes | Yes | Yes | Yes | Yes |
| Day of Week<br>Fixed Effect | Yes | Yes | Yes | Yes | Yes | Yes | Yes |
| Individual<br>Fixed Effect | Yes | Yes | Yes | Yes | Yes | Yes | Yes |
| Log likelihood /<br>R-squared | -621877.48 | -621726.60 | -621849.09 | -621744.97 | 0.522 | 0.522 | 0.523 |
| Observations | 1,279,077 | 1,279,077 | 1,279,077 | 1,279,077 | 1,279,077 | 1,279,077 | 1,279,077 |

$* p < 0.10, ** p < 0.05, *** p < 0.01, **** p < 0.001.$

**Table A6 – Virginia Beach**

| City = Virginia Beach<br>Tr. = National Emergency | (1)<br>Main Effect<br>(Logit) | (2)<br>Interaction<br>w/ Income<br>(Logit) | (3)<br>Interaction<br>w/ Gender<br>(Logit) | (4)<br>Interaction<br>w/ Race<br>(Logit) | (5)<br>Interaction<br>w/ Income<br>(LPM) | (6)<br>Interaction<br>w/ Gender<br>(LPM) | (7)<br>Interaction<br>w/ Race<br>(LPM) |
|---|---|---|---|---|---|---|---|
| Treat | -0.085****<br>(0.009) | | | | | | |
| Treat ×<br>Income less than 60K | | 0.474****<br>(0.063) | | | 0.047****<br>(0.004) | | |
| Treat ×<br>Income 60K – 100K | | 0.796****<br>(0.085) | | | -0.004<br>(0.005) | | |
| Treat ×<br>Income 100K – 150K | | 0.247***<br>(0.078) | | | 0.037****<br>(0.006) | | |
| Treat ×<br>Income 150K – 200K | | 0.661****<br>(0.129) | | | -0.286****<br>(0.010) | | |
| Treat ×<br>Income more than 200K | | -0.575****<br>(0.066) | | | 0.011***<br>(0.004) | | |
| Treat ×<br>Female | | | -0.212****<br>(0.023) | | | -0.023****<br>(0.002) | |
| Treat ×<br>Male | | | 0.246****<br>(0.041) | | | 0.079****<br>(0.004) | |
| Treat ×<br>Race Others | | | | 0.296****<br>(0.084) | | | 0.047****<br>(0.004) |
| Treat ×<br>Race White | | | | -0.521****<br>(0.087) | | | -0.059****<br>(0.004) |
| Treat ×<br>Race Black | | | | -0.032<br>(0.097) | | | 0.070****<br>(0.005) |
| Treat ×<br>Race Asian | | | | -0.265**<br>(0.117) | | | -0.017**<br>(0.008) |
| Treat ×<br>Race Native | | | | -5.026****<br>(0.691) | | | -1.067****<br>(0.070) |
| Mobile App Usage Controlled | Yes | Yes | Yes | Yes | Yes | Yes | Yes |
| Week Fixed Effect | Yes | Yes | Yes | Yes | Yes | Yes | Yes |
| Day of Week Fixed Effect | Yes | Yes | Yes | Yes | Yes | Yes | Yes |
| Individual Fixed Effect | Yes | Yes | Yes | Yes | Yes | Yes | Yes |
| Log likelihood / R-squared | -656051.69 | -655970.38 | -656033.72 | -655828.52 | 0.475 | 0.474 | 0.475 |
| Observations | 1,239,244 | 1,239,244 | 1,239,244 | 1,239,244 | 1,239,244 | 1,239,244 | 1,239,244 |

* $p < 0.10$, ** $p < 0.05$, *** $p < 0.01$, **** $p < 0.001$.

**Table A7 – SFO**

| City = SFO<br>Tr. = National Emergency | (1)<br>Main Effect<br>(Logit) | (2)<br>Interaction<br>w/ Income<br>(Logit) | (3)<br>Interaction<br>w/ Gender<br>(Logit) | (4)<br>Interaction<br>w/ Race<br>(Logit) | (5)<br>Interaction<br>w/ Income<br>(LPM) | (6)<br>Interaction<br>w/ Gender<br>(LPM) | (7)<br>Interaction<br>w/ Race<br>(LPM) |
|---|---|---|---|---|---|---|---|
| Treat | -0.104****<br>(0.010) | | | | | | |
| Treat ×<br>Income less than 60K | | 0.220****<br>(0.044) | | | 0.002<br>(0.002) | | |
| Treat ×<br>Income 60K – 100K | | 0.195****<br>(0.055) | | | 0.062****<br>(0.004) | | |
| Treat ×<br>Income 100K – 150K | | 0.244***<br>(0.073) | | | -0.016****<br>(0.005) | | |
| Treat ×<br>Income 150K – 200K | | 0.411****<br>(0.106) | | | 0.029****<br>(0.007) | | |
| Treat ×<br>Income more than 200K | | -0.311****<br>(0.042) | | | -0.040****<br>(0.002) | | |
| Treat ×<br>Female | | | -0.108***<br>(0.035) | | | -0.011****<br>(0.002) | |
| Treat ×<br>Male | | | 0.009<br>(0.067) | | | -0.035****<br>(0.004) | |
| Treat ×<br>Race Others | | | | 0.226****<br>(0.040) | | | 0.009****<br>(0.002) |
| Treat ×<br>Race White | | | | -0.416****<br>(0.046) | | | -0.060****<br>(0.002) |
| Treat ×<br>Race Black | | | | -0.086<br>(0.068) | | | 0.044****<br>(0.004) |
| Treat ×<br>Race Asian | | | | -0.448****<br>(0.046) | | | -0.058****<br>(0.002) |
| Treat ×<br>Race Native | | | | -1.374****<br>(0.368) | | | 0.093**<br>(0.038) |
| Mobile App Usage<br>Controlled | Yes | Yes | Yes | Yes | Yes | Yes | Yes |
| Week<br>Fixed Effect | Yes | Yes | Yes | Yes | Yes | Yes | Yes |
| Day of Week<br>Fixed Effect | Yes | Yes | Yes | Yes | Yes | Yes | Yes |
| Individual<br>Fixed Effect | Yes | Yes | Yes | Yes | Yes | Yes | Yes |
| Log likelihood /<br>R-squared | -715792.68 | -715779.29 | -715792.67 | -715668.89 | 0.496 | 0.496 | 0.496 |
| Observations | 1,653,110 | 1,653,110 | 1,653,110 | 1,653,110 | 1,653,110 | 1,653,110 | 1,653,110 |

* $p < 0.10$, ** $p < 0.05$, *** $p < 0.01$, **** $p < 0.001$.

**Table A8 – Jacksonville**

| City = Jacksonville<br>Tr. = National Emergency | (1)<br>Main Effect<br>(Logit) | (2)<br>Interaction<br>w/ Income<br>(Logit) | (3)<br>Interaction<br>w/ Gender<br>(Logit) | (4)<br>Interaction<br>w/ Race<br>(Logit) | (5)<br>Interaction<br>w/ Income<br>(LPM) | (6)<br>Interaction<br>w/ Gender<br>(LPM) | (7)<br>Interaction<br>w/ Race<br>(LPM) |
|---|---|---|---|---|---|---|---|
| Treat | -0.089****<br>(0.009) | | | | | | |
| Treat ×<br>Income less than 60K | | -0.302****<br>(0.083) | | | 0.037****<br>(0.007) | | |
| Treat ×<br>Income 60K – 100K | | -0.021<br>(0.091) | | | 0.122****<br>(0.008) | | |
| Treat ×<br>Income 100K – 150K | | -0.676****<br>(0.113) | | | -0.126****<br>(0.011) | | |
| Treat ×<br>Income 150K – 200K | | -1.009****<br>(0.177) | | | 0.015<br>(0.017) | | |
| Treat ×<br>Income more than 200K | | 0.223***<br>(0.081) | | | -0.018**<br>(0.007) | | |
| Treat ×<br>Female | | | 0.198****<br>(0.047) | | | 0.061****<br>(0.004) | |
| Treat ×<br>Male | | | -0.592****<br>(0.094) | | | -0.095****<br>(0.009) | |
| Treat ×<br>Race Others | | | | -0.257**<br>(0.111) | | | -0.052****<br>(0.010) |
| Treat ×<br>Race White | | | | 0.152<br>(0.116) | | | 0.065****<br>(0.010) |
| Treat ×<br>Race Black | | | | 0.043<br>(0.119) | | | 0.059****<br>(0.010) |
| Treat ×<br>Race Asian | | | | 1.294****<br>(0.167) | | | 0.251****<br>(0.016) |
| Treat ×<br>Race Native | | | | -4.966****<br>(0.826) | | | -0.745****<br>(0.085) |
| Mobile App Usage Controlled | Yes | Yes | Yes | Yes | Yes | Yes | Yes |
| Week Fixed Effect | Yes | Yes | Yes | Yes | Yes | Yes | Yes |
| Day of Week Fixed Effect | Yes | Yes | Yes | Yes | Yes | Yes | Yes |
| Individual Fixed Effect | Yes | Yes | Yes | Yes | Yes | Yes | Yes |
| Log likelihood / R-squared | -595232.56 | -595155.02 | -595213.21 | -595134.94 | 0.511 | 0.511 | 0.511 |
| Observations | 1,210,655 | 1,210,655 | 1,210,655 | 1,210,655 | 1,210,655 | 1,210,655 | 1,210,655 |

* $p < 0.10$, ** $p < 0.05$, *** $p < 0.01$, **** $p < 0.001$.

**Table A9 – New Orleans**

| City = New Orleans<br>Tr. = National Emergency | (1)<br>Main Effect<br>(Logit) | (2)<br>Interaction w/ Income<br>(Logit) | (3)<br>Interaction w/ Gender<br>(Logit) | (4)<br>Interaction w/ Race<br>(Logit) | (5)<br>Interaction w/ Income<br>(LPM) | (6)<br>Interaction w/ Gender<br>(LPM) | (7)<br>Interaction w/ Race<br>(LPM) |
|---|---|---|---|---|---|---|---|
| Treat | 0.175****<br>(0.012) | | | | | | |
| Treat × Income less than 60K | | -0.662****<br>(0.093) | | | -0.284****<br>(0.033) | | |
| Treat × Income 60K – 100K | | -0.019<br>(0.111) | | | 0.287****<br>(0.059) | | |
| Treat × Income 100K – 150K | | -1.317****<br>(0.139) | | | -0.884****<br>(0.101) | | |
| Treat × Income 150K – 200K | | 0.322<br>(0.207) | | | 0.835****<br>(0.135) | | |
| Treat × Income more than 200K | | 0.697****<br>(0.094) | | | 0.334****<br>(0.033) | | |
| Treat × Female | | | 0.155****<br>(0.031) | | | -0.151****<br>(0.004) | |
| Treat × Male | | | 0.040<br>(0.057) | | | 0.130****<br>(0.002) | |
| Treat × Race Others | | | | -0.892****<br>(0.081) | | | 0.066****<br>(0.005) |
| Treat × Race White | | | | 1.201****<br>(0.084) | | | -0.023****<br>(0.005) |
| Treat × Race Black | | | | 1.000****<br>(0.084) | | | -0.001<br>(0.006) |
| Treat × Race Asian | | | | 1.501****<br>(0.171) | | | -0.178****<br>(0.011) |
| Treat × Race Native | | | | -1.102***<br>(0.349) | | | 0.550****<br>(0.042) |
| Mobile App Usage Controlled | Yes | Yes | Yes | Yes | Yes | Yes | Yes |
| Week Fixed Effect | Yes | Yes | Yes | Yes | Yes | Yes | Yes |
| Day of Week Fixed Effect | Yes | Yes | Yes | Yes | Yes | Yes | Yes |
| Individual Fixed Effect | Yes | Yes | Yes | Yes | Yes | Yes | Yes |
| Log likelihood / R-squared | -593304.67 | -593102.40 | -593304.42 | -593132.71 | 0.516 | 0.516 | 0.516 |
| Observations | 1,486,512 | 1,486,512 | 1,486,512 | 1,486,512 | 1,486,512 | 1,486,512 | 1,486,512 |

* $p < 0.10$, ** $p < 0.05$, *** $p < 0.01$, **** $p < 0.001$.

**Table A10 – Omaha**

| City = Omaha<br>Tr. = National Emergency | (1)<br>Main Effect<br>(Logit) | (2)<br>Interaction<br>w/ Income<br>(Logit) | (3)<br>Interaction<br>w/ Gender<br>(Logit) | (4)<br>Interaction<br>w/ Race<br>(Logit) | (5)<br>Interaction<br>w/ Income<br>(LPM) | (6)<br>Interaction<br>w/ Gender<br>(LPM) | (7)<br>Interaction<br>w/ Race<br>(LPM) |
|---|---|---|---|---|---|---|---|
| Treat | -0.112****<br>(0.008) | | | | | | |
| Treat ×<br>Income less than 60K | | -0.841****<br>(0.082) | | | -0.116****<br>(0.008) | | |
| Treat ×<br>Income 60K – 100K | | -0.751****<br>(0.088) | | | -0.083****<br>(0.009) | | |
| Treat ×<br>Income 100K – 150K | | -0.667****<br>(0.103) | | | -0.230****<br>(0.011) | | |
| Treat ×<br>Income 150K – 200K | | -1.474****<br>(0.156) | | | -0.331****<br>(0.017) | | |
| Treat ×<br>Income more than 200K | | 0.681****<br>(0.082) | | | 0.145****<br>(0.008) | | |
| Treat ×<br>Female | | | -0.169****<br>(0.038) | | | 0.001<br>(0.004) | |
| Treat ×<br>Male | | | 0.114<br>(0.076) | | | 0.029***<br>(0.009) | |
| Treat ×<br>Race Others | | | | -0.731****<br>(0.067) | | | -0.122****<br>(0.008) |
| Treat ×<br>Race White | | | | 0.676****<br>(0.070) | | | 0.144****<br>(0.008) |
| Treat ×<br>Race Black | | | | 0.318****<br>(0.080) | | | 0.074****<br>(0.009) |
| Treat ×<br>Race Asian | | | | 1.012****<br>(0.095) | | | 0.341****<br>(0.012) |
| Treat ×<br>Race Native | | | | -1.322****<br>(0.375) | | | 0.063<br>(0.045) |
| Mobile App Usage Controlled | Yes | Yes | Yes | Yes | Yes | Yes | Yes |
| Week Fixed Effect | Yes | Yes | Yes | Yes | Yes | Yes | Yes |
| Day of Week Fixed Effect | Yes | Yes | Yes | Yes | Yes | Yes | Yes |
| Individual Fixed Effect | Yes | Yes | Yes | Yes | Yes | Yes | Yes |
| Log likelihood /<br>R-squared | -717022.55 | -716956.06 | -717021.42 | -716885.82 | 0.436 | 0.436 | 0.437 |
| Observations | 1,331,130 | 1,331,130 | 1,331,130 | 1,331,130 | 1,331,130 | 1,331,130 | 1,331,130 |

* $p < 0.10$, ** $p < 0.05$, *** $p < 0.01$, **** $p < 0.001$.

**Table A11 – NYC**

| City = NYC<br>Tr. = National Emergency | (1)<br>Main Effect<br>(Logit) | (2)<br>Interaction<br>w/ Income<br>(Logit) | (3)<br>Interaction<br>w/ Gender<br>(Logit) | (4)<br>Interaction<br>w/ Race<br>(Logit) | (5)<br>Interaction<br>w/ Income<br>(LPM) | (6)<br>Interaction<br>w/ Gender<br>(LPM) | (7)<br>Interaction<br>w/ Race<br>(LPM) |
|---|---|---|---|---|---|---|---|
| Treat | 0.133****<br>(0.012) | | | | | | |
| Treat ×<br>Income less than 60K | | -0.579****<br>(0.026) | | | -0.144****<br>(0.002) | | |
| Treat ×<br>Income 60K – 100K | | -0.804****<br>(0.046) | | | -0.122****<br>(0.005) | | |
| Treat ×<br>Income 100K – 150K | | -0.496****<br>(0.058) | | | -0.087****<br>(0.006) | | |
| Treat ×<br>Income 150K – 200K | | 0.494****<br>(0.076) | | | 0.093****<br>(0.008) | | |
| Treat ×<br>Income more than 200K | | 0.573****<br>(0.024) | | | 0.158****<br>(0.002) | | |
| Treat ×<br>Female | | | 0.023<br>(0.022) | | | 0.043****<br>(0.002) | |
| Treat ×<br>Male | | | 0.232****<br>(0.039) | | | 0.048****<br>(0.004) | |
| Treat ×<br>Race Others | | | | 0.007<br>(0.025) | | | 0.068****<br>(0.003) |
| Treat ×<br>Race White | | | | 0.183****<br>(0.026) | | | 0.008****<br>(0.003) |
| Treat ×<br>Race Black | | | | 0.030<br>(0.029) | | | -0.072****<br>(0.003) |
| Treat ×<br>Race Asian | | | | 0.122****<br>(0.033) | | | 0.036****<br>(0.003) |
| Treat ×<br>Race Native | | | | 0.999**<br>(0.402) | | | 0.178****<br>(0.043) |
| Mobile App Usage<br>Controlled | Yes | Yes | Yes | Yes | Yes | Yes | Yes |
| Week<br>Fixed Effect | Yes | Yes | Yes | Yes | Yes | Yes | Yes |
| Day of Week<br>Fixed Effect | Yes | Yes | Yes | Yes | Yes | Yes | Yes |
| Individual<br>Fixed Effect | Yes | Yes | Yes | Yes | Yes | Yes | Yes |
| Log likelihood / R-squared | -657206.31 | -656664.26 | -657189.48 | -657156.01 | 0.476 | 0.475 | 0.476 |
| Observations | 1,310,678 | 1,310,678 | 1,310,678 | 1,310,678 | 1,310,678 | 1,310,678 | 1,310,678 |

* $p < 0.10$, ** $p < 0.05$, *** $p < 0.01$, **** $p < 0.001$.

**Table A12 – Pittsburgh**

| City = Pittsburgh<br>Tr. = National Emergency | (1)<br>Main Effect<br>(Logit) | (2)<br>Interaction<br>w/ Income<br>(Logit) | (3)<br>Interaction<br>w/ Gender<br>(Logit) | (4)<br>Interaction<br>w/ Race<br>(Logit) | (5)<br>Interaction<br>w/ Income<br>(LPM) | (6)<br>Interaction<br>w/ Gender<br>(LPM) | (7)<br>Interaction<br>w/ Race<br>(LPM) |
|---|---|---|---|---|---|---|---|
| Treat | 0.136****<br>(0.008) | | | | | | |
| Treat ×<br>Income less than 60K | | -0.081*<br>(0.045) | | | -0.003<br>(0.004) | | |
| Treat ×<br>Income 60K – 100K | | -0.256****<br>(0.061) | | | -0.049****<br>(0.006) | | |
| Treat ×<br>Income 100K – 150K | | -0.237****<br>(0.065) | | | -0.016**<br>(0.006) | | |
| Treat ×<br>Income 150K – 200K | | 0.884****<br>(0.115) | | | 0.005<br>(0.011) | | |
| Treat ×<br>Income more than 200K | | 0.145****<br>(0.044) | | | 0.057****<br>(0.004) | | |
| Treat ×<br>Female | | | 0.230****<br>(0.031) | | | 0.024****<br>(0.003) | |
| Treat ×<br>Male | | | -0.352****<br>(0.062) | | | 0.036****<br>(0.006) | |
| Treat ×<br>Race Others | | | | -0.041<br>(0.076) | | | 0.065****<br>(0.008) |
| Treat ×<br>Race White | | | | 0.087<br>(0.077) | | | -0.026****<br>(0.008) |
| Treat ×<br>Race Black | | | | -0.229***<br>(0.087) | | | -0.075****<br>(0.009) |
| Treat ×<br>Race Asian | | | | 1.380****<br>(0.126) | | | 0.205****<br>(0.012) |
| Treat ×<br>Race Native | | | | 1.450<br>(1.133) | | | -0.534****<br>(0.097) |
| Mobile App Usage Controlled | Yes | Yes | Yes | Yes | Yes | Yes | Yes |
| Week Fixed Effect | Yes | Yes | Yes | Yes | Yes | Yes | Yes |
| Day of Week Fixed Effect | Yes | Yes | Yes | Yes | Yes | Yes | Yes |
| Individual Fixed Effect | Yes | Yes | Yes | Yes | Yes | Yes | Yes |
| Log likelihood / R-squared | -599325.79 | -596372.07 | -596424.72 | -596308.86 | 0.482 | 0.482 | 0.483 |
| Observations | 2,063,683 | 2,063,683 | 2,063,683 | 2,063,683 | 2,063,683 | 2,063,683 | 2,063,683 |

* $p < 0.10$, ** $p < 0.05$, *** $p < 0.01$, **** $p < 0.001$.

**Table A13 – Oklahoma City**

| City = Oklahoma City<br>Tr. = National Emergency | (1)<br>Main Effect<br>(Logit) | (2)<br>Interaction<br>w/ Income<br>(Logit) | (3)<br>Interaction<br>w/ Gender<br>(Logit) | (4)<br>Interaction<br>w/ Race<br>(Logit) | (5)<br>Interaction<br>w/ Income<br>(LPM) | (6)<br>Interaction<br>w/ Gender<br>(LPM) | (7)<br>Interaction<br>w/ Race<br>(LPM) |
|---|---|---|---|---|---|---|---|
| Treat | -0.029***<br>(0.009) | | | | | | |
| Treat ×<br>Income less than 60K | | -0.113***<br>(0.043) | | | -0.008*<br>(0.004) | | |
| Treat ×<br>Income 60K – 100K | | -0.382****<br>(0.055) | | | -0.058****<br>(0.005) | | |
| Treat ×<br>Income 100K – 150K | | -0.607****<br>(0.077) | | | -0.203****<br>(0.008) | | |
| Treat ×<br>Income 150K – 200K | | -0.403****<br>(0.119) | | | -0.018<br>(0.013) | | |
| Treat ×<br>Income more than 200K | | 0.214****<br>(0.042) | | | 0.075****<br>(0.004) | | |
| Treat ×<br>Female | | | -0.160****<br>(0.026) | | | -0.015****<br>(0.003) | |
| Treat ×<br>Male | | | 0.261****<br>(0.049) | | | 0.092****<br>(0.005) | |
| Treat ×<br>Race Others | | | | 0.172****<br>(0.037) | | | 0.073****<br>(0.003) |
| Treat ×<br>Race White | | | | -0.190****<br>(0.041) | | | -0.046****<br>(0.004) |
| Treat ×<br>Race Black | | | | -0.345****<br>(0.049) | | | -0.046****<br>(0.005) |
| Treat ×<br>Race Asian | | | | -0.123<br>(0.097) | | | -0.079****<br>(0.010) |
| Treat ×<br>Race Native | | | | -0.682****<br>(0.126) | | | -0.045***<br>(0.015) |
| Mobile App Usage Controlled | Yes | Yes | Yes | Yes | Yes | Yes | Yes |
| Week Fixed Effect | Yes | Yes | Yes | Yes | Yes | Yes | Yes |
| Day of Week Fixed Effect | Yes | Yes | Yes | Yes | Yes | Yes | Yes |
| Individual Fixed Effect | Yes | Yes | Yes | Yes | Yes | Yes | Yes |
| Log likelihood / R-squared | -643904.68 | -643785.43 | -643890.65 | -643872.94 | 0.476 | 0.462 | 0.462 |
| Observations | 1,206,577 | 1,206,577 | 1,206,577 | 1,206,577 | 1,206,577 | 1,206,577 | 1,206,577 |

* $p < 0.10$, ** $p < 0.05$, *** $p < 0.01$, **** $p < 0.001$.

**Table A14 – Philadelphia**

| City = Philadelphia<br>Tr. = National Emergency | (1)<br>Main Effect<br>(Logit) | (2)<br>Interaction<br>w/ Income<br>(Logit) | (3)<br>Interaction<br>w/ Gender<br>(Logit) | (4)<br>Interaction<br>w/ Race<br>(Logit) | (5)<br>Interaction<br>w/ Income<br>(LPM) | (6)<br>Interaction<br>w/ Gender<br>(LPM) | (7)<br>Interaction<br>w/ Race<br>(LPM) |
|---|---|---|---|---|---|---|---|
| Treat | -0.028****<br>(0.010) | | | | | | |
| Treat ×<br>Income less than 60K | | -0.663****<br>(0.020) | | | -0.149****<br>(0.002) | | |
| Treat ×<br>Income 60K – 100K | | -0.429****<br>(0.045) | | | -0.103****<br>(0.005) | | |
| Treat ×<br>Income 100K – 150K | | -0.517****<br>(0.051) | | | -0.065****<br>(0.006) | | |
| Treat ×<br>Income 150K – 200K | | -0.197****<br>(0.080) | | | -0.029****<br>(0.009) | | |
| Treat ×<br>Income more than 200K | | 0.446****<br>(0.018) | | | 0.136****<br>(0.002) | | |
| Treat ×<br>Female | | | 0.269****<br>(0.017) | | | 0.102****<br>(0.002) | |
| Treat ×<br>Male | | | -0.667****<br>(0.031) | | | -0.149****<br>(0.003) | |
| Treat ×<br>Race Others | | | | 0.320****<br>(0.018) | | | 0.113****<br>(0.002) |
| Treat ×<br>Race White | | | | -0.363****<br>(0.018) | | | -0.077****<br>(0.002) |
| Treat ×<br>Race Black | | | | -0.565****<br>(0.022) | | | -0.139****<br>(0.002) |
| Treat ×<br>Race Asian | | | | -0.141***<br>(0.052) | | | -0.023****<br>(0.005) |
| Treat ×<br>Race Native | | | | -4.645****<br>(0.512) | | | -0.903****<br>(0.055) |
| Mobile App Usage<br>Controlled | Yes | Yes | Yes | Yes | Yes | Yes | Yes |
| Week<br>Fixed Effect | Yes | Yes | Yes | Yes | Yes | Yes | Yes |
| Day of Week<br>Fixed Effect | Yes | Yes | Yes | Yes | Yes | Yes | Yes |
| Individual<br>Fixed Effect | Yes | Yes | Yes | Yes | Yes | Yes | Yes |
| Log likelihood / R-squared | -682891.69 | -682257.77 | -682666.95 | -682539.46 | 0.453 | 0.452 | 0.452 |
| Observations | 1,319,188 | 1,319,188 | 1,319,188 | 1,319,188 | 1,319,188 | 1,319,188 | 1,319,188 |

* $p < 0.10$, ** $p < 0.05$, *** $p < 0.01$, **** $p < 0.001$.

**Table A15 – Austin**

| City = Austin<br>Tr. = National Emergency | (1)<br>Main Effect<br>(Logit) | (2)<br>Interaction<br>w/ Income<br>(Logit) | (3)<br>Interaction<br>w/ Gender<br>(Logit) | (4)<br>Interaction<br>w/ Race<br>(Logit) | (5)<br>Interaction<br>w/ Income<br>(LPM) | (6)<br>Interaction<br>w/ Gender<br>(LPM) | (7)<br>Interaction<br>w/ Race<br>(LPM) |
|---|---|---|---|---|---|---|---|
| Treat | -0.026<br>(0.010) | | | | | | |
| Treat ×<br>Income less than 60K | | -0.116****<br>(0.023) | | | -0.011****<br>(0.002) | | |
| Treat ×<br>Income 60K – 100K | | 0.116****<br>(0.041) | | | 0.041****<br>(0.004) | | |
| Treat ×<br>Income 100K – 150K | | -0.659****<br>(0.057) | | | -0.177****<br>(0.006) | | |
| Treat ×<br>Income 150K – 200K | | -0.410****<br>(0.089) | | | -0.088****<br>(0.009) | | |
| Treat ×<br>Income more than 200K | | 0.117****<br>(0.020) | | | 0.059****<br>(0.002) | | |
| Treat ×<br>Female | | | -0.060***<br>(0.021) | | | 0.020****<br>(0.002) | |
| Treat ×<br>Male | | | 0.067*<br>(0.037) | | | 0.026****<br>(0.004) | |
| Treat ×<br>Race Others | | | | 0.090****<br>(0.025) | | | 0.064****<br>(0.002) |
| Treat ×<br>Race White | | | | -0.139****<br>(0.026) | | | -0.038****<br>(0.003) |
| Treat ×<br>Race Black | | | | 0.136**<br>(0.056) | | | 0.069****<br>(0.006) |
| Treat ×<br>Race Asian | | | | -0.469****<br>(0.069) | | | -0.143****<br>(0.007) |
| Treat ×<br>Race Native | | | | 0.091<br>(0.404) | | | -0.079**<br>(0.046) |
| Mobile App Usage<br>Controlled | Yes | Yes | Yes | Yes | Yes | Yes | Yes |
| Week<br>Fixed Effect | Yes | Yes | Yes | Yes | Yes | Yes | Yes |
| Day of Week<br>Fixed Effect | Yes | Yes | Yes | Yes | Yes | Yes | Yes |
| Individual<br>Fixed Effect | Yes | Yes | Yes | Yes | Yes | Yes | Yes |
| Log likelihood / R-squared | -651003.54 | -650891.85 | -651001.96 | -650956.85 | 0.480 | 0.479 | 0.480 |
| Observations | 1,260,615 | 1,260,615 | 1,260,615 | 1,260,615 | 1,260,615 | 1,260,615 | 1,260,615 |

\* $p < 0.10$, ** $p < 0.05$, *** $p < 0.01$, **** $p < 0.001$.

**Table A16 – Seattle**

| City = Seattle<br>Tr. = National Emergency | (1)<br>Main Effect<br>(Logit) | (2)<br>Interaction<br>w/ Income<br>(Logit) | (3)<br>Interaction<br>w/ Gender<br>(Logit) | (4)<br>Interaction<br>w/ Race<br>(Logit) | (5)<br>Interaction<br>w/ Income<br>(LPM) | (6)<br>Interaction<br>w/ Gender<br>(LPM) | (7)<br>Interaction<br>w/ Race<br>(LPM) |
|---|---|---|---|---|---|---|---|
| Treat | -0.286****<br>(0.012) | | | | | | |
| Treat ×<br>Income less than 60K | | 0.298****<br>(0.058) | | | 0.070****<br>(0.004) | | |
| Treat ×<br>Income 60K – 100K | | 0.319****<br>(0.070) | | | 0.024****<br>(0.005) | | |
| Treat ×<br>Income 100K – 150K | | 0.127<br>(0.082) | | | 0.006****<br>(0.006) | | |
| Treat ×<br>Income 150K – 200K | | 0.201<br>(0.128) | | | -0.072****<br>(0.010) | | |
| Treat ×<br>Income more than 200K | | -0.538****<br>(0.057) | | | -0.042****<br>(0.004) | | |
| Treat ×<br>Female | | | -0.239****<br>(0.041) | | | -0.015****<br>(0.003) | |
| Treat ×<br>Male | | | -0.093<br>(0.080) | | | 0.017****<br>(0.006) | |
| Treat ×<br>Race Others | | | | -0.237****<br>(0.050) | | | 0.001<br>(0.004) |
| Treat ×<br>Race White | | | | -0.076<br>(0.054) | | | -0.010**<br>(0.004) |
| Treat ×<br>Race Black | | | | 0.497****<br>(0.079) | | | 0.088****<br>(0.006) |
| Treat ×<br>Race Asian | | | | -0.253****<br>(0.070) | | | -0.065****<br>(0.005) |
| Treat ×<br>Race Native | | | | -0.120<br>(0.215) | | | 0.074****<br>(0.016) |
| Mobile App Usage Controlled | Yes | Yes | Yes | Yes | Yes | Yes | Yes |
| Week Fixed Effect | Yes | Yes | Yes | Yes | Yes | Yes | Yes |
| Day of Week Fixed Effect | Yes | Yes | Yes | Yes | Yes | Yes | Yes |
| Individual Fixed Effect | Yes | Yes | Yes | Yes | Yes | Yes | Yes |
| Log likelihood / R-squared | -518579.67 | -518549.17 | -518579.00 | -518503.34 | 0.559 | 0.559 | 0.559 |
| Observations | 989,083 | 989,083 | 989,083 | 989,083 | 989,083 | 989,083 | 989,083 |

* $p < 0.10$, ** $p < 0.05$, *** $p < 0.01$, **** $p < 0.001$.

**Table A17 – Arlington**

| City = Arlington<br>Tr. = National Emergency | (1)<br>Main Effect<br>(Logit) | (2)<br>Interaction<br>w/ Income<br>(Logit) | (3)<br>Interaction<br>w/ Gender<br>(Logit) | (4)<br>Interaction<br>w/ Race<br>(Logit) | (5)<br>Interaction<br>w/ Income<br>(LPM) | (6)<br>Interaction<br>w/ Gender<br>(LPM) | (7)<br>Interaction<br>w/ Race<br>(LPM) |
|---|---|---|---|---|---|---|---|
| Treat | 0.066****<br>(0.009) | | | | | | |
| Treat ×<br>Income less than 60K | | -0.385****<br>(0.117) | | | 0.005<br>(0.015) | | |
| Treat ×<br>Income 60K – 100K | | -0.382****<br>(0.105) | | | 0.090****<br>(0.014) | | |
| Treat ×<br>Income 100K – 150K | | -0.572****<br>(0.162) | | | -0.064***<br>(0.021) | | |
| Treat ×<br>Income 150K – 200K | | -0.285*<br>(0.161) | | | -0.201****<br>(0.022) | | |
| Treat ×<br>Income more than 200K | | 0.460****<br>(0.117) | | | 0.053****<br>(0.015) | | |
| Treat ×<br>Female | | | -0.439****<br>(0.051) | | | -0.075****<br>(0.006) | |
| Treat ×<br>Male | | | 1.029****<br>(0.102) | | | 0.265****<br>(0.013) | |
| Treat ×<br>Race Others | | | | 0.133*<br>(0.073) | | | 0.130****<br>(0.010) |
| Treat ×<br>Race White | | | | -0.051<br>(0.079) | | | -0.086****<br>(0.011) |
| Treat ×<br>Race Black | | | | -0.232***<br>(0.079) | | | -0.089****<br>(0.011) |
| Treat ×<br>Race Asian | | | | 0.225**<br>(0.104) | | | -0.044****<br>(0.014) |
| Treat ×<br>Race Native | | | | 1.675****<br>(0.373) | | | 0.433****<br>(0.052) |
| Mobile App Usage Controlled | Yes | Yes | Yes | Yes | Yes | Yes | Yes |
| Week Fixed Effect | Yes | Yes | Yes | Yes | Yes | Yes | Yes |
| Day of Week Fixed Effect | Yes | Yes | Yes | Yes | Yes | Yes | Yes |
| Individual Fixed Effect | Yes | Yes | Yes | Yes | Yes | Yes | Yes |
| Log likelihood / R-squared | -757979.10 | -757968.40 | -757928.76 | -757933.83 | 0.382 | 0.382 | 0.382 |
| Observations | 1,402,332 | 1,402,332 | 1,402,332 | 1,402,332 | 1,402,332 | 1,402,332 | 1,402,332 |

* $p < 0.10$, ** $p < 0.05$, *** $p < 0.01$, **** $p < 0.001$.

**Table A18 – Phoenix**

| City = Phoenix<br>Tr. = National Emergency | (1)<br>Main Effect<br>(Logit) | (2)<br>Interaction<br>w/ Income<br>(Logit) | (3)<br>Interaction<br>w/ Gender<br>(Logit) | (4)<br>Interaction<br>w/ Race<br>(Logit) | (5)<br>Interaction<br>w/ Income<br>(LPM) | (6)<br>Interaction<br>w/ Gender<br>(LPM) | (7)<br>Interaction<br>w/ Race<br>(LPM) |
|---|---|---|---|---|---|---|---|
| Treat | -0.149****<br>(0.011) | | | | | | |
| Treat ×<br>Income less than 60K | | -0.262****<br>(0.030) | | | -0.005**<br>(0.002) | | |
| Treat ×<br>Income 60K – 100K | | -0.878****<br>(0.052) | | | -0.108****<br>(0.004) | | |
| Treat ×<br>Income 100K – 150K | | -0.100<br>(0.071) | | | -0.055****<br>(0.005) | | |
| Treat ×<br>Income 150K – 200K | | -1.312****<br>(0.116) | | | -0.201****<br>(0.008) | | |
| Treat ×<br>Income more than 200K | | 0.251****<br>(0.029) | | | 0.056****<br>(0.002) | | |
| Treat ×<br>Female | | | 0.047*<br>(0.026) | | | 0.027****<br>(0.002) | |
| Treat ×<br>Male | | | -0.404****<br>(0.048) | | | -0.031****<br>(0.004) | |
| Treat ×<br>Race Others | | | | 0.199****<br>(0.029) | | | 0.053****<br>(0.002) |
| Treat ×<br>Race White | | | | -0.393****<br>(0.031) | | | -0.051****<br>(0.002) |
| Treat ×<br>Race Black | | | | -0.958****<br>(0.085) | | | -0.076****<br>(0.006) |
| Treat ×<br>Race Asian | | | | -0.690****<br>(0.098) | | | -0.122****<br>(0.007) |
| Treat ×<br>Race Native | | | | 0.160****<br>(0.040) | | | 0.112****<br>(0.004) |
| Mobile App Usage Controlled | Yes | Yes | Yes | Yes | Yes | Yes | Yes |
| Week Fixed Effect | Yes | Yes | Yes | Yes | Yes | Yes | Yes |
| Day of Week Fixed Effect | Yes | Yes | Yes | Yes | Yes | Yes | Yes |
| Individual Fixed Effect | Yes | Yes | Yes | Yes | Yes | Yes | Yes |
| Log likelihood / R-squared | -492242.38 | -492019.28 | -492207.64 | -491948.07 | 0.603 | 0.602 | 0.603 |
| Observations | 1,028,752 | 1,028,752 | 1,028,752 | 1,028,752 | 1,028,752 | 1,028,752 | 1,028,752 |

* $p < 0.10$, ** $p < 0.05$, *** $p < 0.01$, **** $p < 0.001$.

**Table A19 – Nashville**

| City = Nashville<br>Tr. = National Emergency | (1)<br>Main Effect<br>(Logit) | (2)<br>Interaction<br>w/ Income<br>(Logit) | (3)<br>Interaction<br>w/ Gender<br>(Logit) | (4)<br>Interaction<br>w/ Race<br>(Logit) | (5)<br>Interaction<br>w/ Income<br>(LPM) | (6)<br>Interaction<br>w/ Gender<br>(LPM) | (7)<br>Interaction<br>w/ Race<br>(LPM) |
|---|---|---|---|---|---|---|---|
| Treat | -0.087****<br>(0.009) | | | | | | |
| Treat ×<br>Income less than 60K | | -0.315****<br>(0.023) | | | -0.011****<br>(0.002) | | |
| Treat ×<br>Income 60K – 100K | | -0.581****<br>(0.042) | | | -0.029****<br>(0.005) | | |
| Treat ×<br>Income 100K – 150K | | 0.123*<br>(0.067) | | | 0.036****<br>(0.007) | | |
| Treat ×<br>Income 150K – 200K | | -1.989****<br>(0.119) | | | -0.363****<br>(0.013) | | |
| Treat ×<br>Income more than 200K | | 0.273****<br>(0.021) | | | 0.049****<br>(0.002) | | |
| Treat ×<br>Female | | | 0.126****<br>(0.020) | | | 0.051****<br>(0.002) | |
| Treat ×<br>Male | | | -0.443****<br>(0.037) | | | -0.054****<br>(0.004) | |
| Treat ×<br>Race Others | | | | 0.219****<br>(0.025) | | | 0.043****<br>(0.002) |
| Treat ×<br>Race White | | | | -0.418****<br>(0.025) | | | -0.034****<br>(0.002) |
| Treat ×<br>Race Black | | | | -0.253****<br>(0.030) | | | 0.005<br>(0.003) |
| Treat ×<br>Race Asian | | | | 1.328****<br>(0.116) | | | 0.235****<br>(0.013) |
| Treat ×<br>Race Native | | | | -3.494****<br>(0.572) | | | -0.906****<br>(0.060) |
| Mobile App Usage Controlled | Yes | Yes | Yes | Yes | Yes | Yes | Yes |
| Week Fixed Effect | Yes | Yes | Yes | Yes | Yes | Yes | Yes |
| Day of Week Fixed Effect | Yes | Yes | Yes | Yes | Yes | Yes | Yes |
| Individual Fixed Effect | Yes | Yes | Yes | Yes | Yes | Yes | Yes |
| Log likelihood / R-squared | -682305.99 | -682052.04 | -682234.92 | -682066.34 | 0.448 | 0.448 | 0.448 |
| Observations | 1,361,080 | 1,361,080 | 1,361,080 | 1,361,080 | 1,361,080 | 1,361,080 | 1,361,080 |

* $p < 0.10$, ** $p < 0.05$, *** $p < 0.01$, **** $p < 0.001$.

**Table A20 – Wichita**

| City = Wichita<br>Tr. = National Emergency | (1)<br>Main Effect<br>(Logit) | (2)<br>Interaction<br>w/ Income<br>(Logit) | (3)<br>Interaction<br>w/ Gender<br>(Logit) | (4)<br>Interaction<br>w/ Race<br>(Logit) | (5)<br>Interaction<br>w/ Income<br>(LPM) | (6)<br>Interaction<br>w/ Gender<br>(LPM) | (7)<br>Interaction<br>w/ Race<br>(LPM) |
|---|---|---|---|---|---|---|---|
| Treat | -0.023****<br>(0.008) | | | | | | |
| Treat ×<br>Income less than 60K | | 1.667****<br>(0.089) | | | -0.377****<br>(0.011) | | |
| Treat ×<br>Income 60K – 100K | | -1.689****<br>(0.090) | | | -0.509****<br>(0.011) | | |
| Treat ×<br>Income 100K – 150K | | -1.931****<br>(0.097) | | | -0.365****<br>(0.013) | | |
| Treat ×<br>Income 150K – 200K | | -1.352****<br>(0.111) | | | -0.750****<br>(0.026) | | |
| Treat ×<br>Income more than 200K | | -3.360****<br>(0.232) | | | 0.431****<br>(0.011) | | |
| Treat ×<br>Female | | | 0.289****<br>(0.037) | | | 0.019****<br>(0.004) | |
| Treat ×<br>Male | | | -0.617****<br>(0.072) | | | 0.022****<br>(0.008) | |
| Treat ×<br>Race Others | | | | -0.724****<br>(0.084) | | | -0.104****<br>(0.010) |
| Treat ×<br>Race White | | | | 0.719****<br>(0.087) | | | 0.135****<br>(0.010) |
| Treat ×<br>Race Black | | | | 0.772****<br>(0.100) | | | 0.188****<br>(0.011) |
| Treat ×<br>Race Asian | | | | 1.056****<br>(0.117) | | | 0.198****<br>(0.013) |
| Treat ×<br>Race Native | | | | 0.586**<br>(0.231) | | | 0.093***<br>(0.027) |
| Mobile App Usage Controlled | Yes | Yes | Yes | Yes | Yes | Yes | Yes |
| Week Fixed Effect | Yes | Yes | Yes | Yes | Yes | Yes | Yes |
| Day of Week Fixed Effect | Yes | Yes | Yes | Yes | Yes | Yes | Yes |
| Individual Fixed Effect | Yes | Yes | Yes | Yes | Yes | Yes | Yes |
| Log likelihood / R-squared | -645866.14 | -645645.75 | -645829.84 | -646011.79 | 0.456 | 0.455 | 0.455 |
| Observations | 1,215,067 | 1,215,067 | 1,215,067 | 1,215,067 | 1,215,067 | 1,215,067 | 1,215,067 |

* $p < 0.10$, ** $p < 0.05$, *** $p < 0.01$, **** $p < 0.001$.

# Appendix B – Per City Results [2]

# Effect of COVID-19 Emergency and Social Distancing on Privacy Concern - Block Level Analyses

### Table B1 – Boston

| City = Boston<br>Tr. = National Emergency | (1)<br>Main Effect | (2)<br>Interaction w/ Daily Contacts |
|---|---|---|
| Treat | -0.562**** <br>(0.026) | -0.615**** <br>(0.028) |
| Treat × Total Daily Contacts |  | 0.063**** <br>(0.012) |
| Total Daily Contacts | 0.310**** <br>(0.007) | 0.289**** <br>(0.008) |
| Daily Travel Distance | 0.228**** <br>(0.010) | 0.223**** <br>(0.010) |
| Daily Avg. Travel Speed | 0.064**** <br>(0.001) | 0.064**** <br>(0.001) |
| Population Controlled | Yes | Yes |
| Block Land Area Controlled | Yes | Yes |
| Population Income Controlled | Yes | Yes |
| Population Gender Controlled | Yes | Yes |
| Population Race Controlled | Yes | Yes |
| Number of Existing Users Controlled | Yes | Yes |
| Number of Opt-in Users Controlled | Yes | Yes |
| Mobile App Usage Controlled | Yes | Yes |
| Week Fixed Effect | Yes | Yes |
| Day of Week Fixed Effect | Yes | Yes |
| Log likelihood | -51128.49 | -51114.74 |
| Observations | 97,308 | 97,308 |

* $p < 0.10$, ** $p < 0.05$, *** $p < 0.01$, **** $p < 0.001$.

**Table B2 – D.C.**

| City = D.C.<br>Tr. = National Emergency | (1)<br>Main Effect | (2)<br>Interaction w/ Daily Contacts |
|---|---|---|
| Treat | -0.262****<br>(0.028) | -0.416****<br>(0.029) |
| Treat ×<br>Total Daily Contacts |  | 0.219****<br>(0.012) |
| Total Daily Contacts | 0.265****<br>(0.006) | 0.206****<br>(0.007) |
| Daily Travel Distance | 0.358****<br>(0.010) | 0.363****<br>(0.010) |
| Daily Avg. Travel Speed | 0.081****<br>(0.001) | 0.081****<br>(0.001) |
| Population<br>Controlled | Yes | Yes |
| Block Land Area<br>Controlled | Yes | Yes |
| Population Income<br>Controlled | Yes | Yes |
| Population Gender<br>Controlled | Yes | Yes |
| Population Race<br>Controlled | Yes | Yes |
| Number of Existing Users<br>Controlled | Yes | Yes |
| Number of Opt-in Users<br>Controlled | Yes | Yes |
| Mobile App Usage<br>Controlled | Yes | Yes |
| Week<br>Fixed Effect | Yes | Yes |
| Day of Week<br>Fixed Effect | Yes | Yes |
| Log likelihood | -43984.82 | -43823.91 |
| Observations | 84,270 | 84,270 |

* $p < 0.10$, ** $p < 0.05$, *** $p < 0.01$, **** $p < 0.001$.

**Table B3 – Baltimore**

| City = Baltimore<br>Tr. = National Emergency | (1)<br>Main Effect | (2)<br>Interaction w/ Daily Contacts |
|---|---|---|
| Treat | -0.337**** | -0.520**** |
|  | (0.027) | (0.029) |
| Treat ×<br>Total Daily Contacts |  | 0.248**** |
|  |  | (0.015) |
| Total Daily Contacts | 0.257**** | 0.169**** |
|  | (0.008) | (0.010) |
| Daily Travel Distance | 0.361**** | 0.345**** |
|  | (0.012) | (0.012) |
| Daily Avg. Travel Speed | 0.087**** | 0.087**** |
|  | (0.001) | (0.001) |
| Population Controlled | Yes | Yes |
| Block Land Area Controlled | Yes | Yes |
| Population Income Controlled | Yes | Yes |
| Population Gender Controlled | Yes | Yes |
| Population Race Controlled | Yes | Yes |
| Number of Existing Users Controlled | Yes | Yes |
| Number of Opt-in Users Controlled | Yes | Yes |
| Mobile App Usage Controlled | Yes | Yes |
| Week Fixed Effect | Yes | Yes |
| Day of Week Fixed Effect | Yes | Yes |
| Log likelihood | -33905.73 | -33764.49 |
| Observations | 71,126 | 71,126 |

\* $p < 0.10$, ** $p < 0.05$, *** $p < 0.01$, **** $p < 0.001$.

## Table B4 – Lexington

| City = Lexington | (1) | (2) |
|---|---|---|
| Tr. = National Emergency | Main Effect | Interaction w/ Daily Contacts |
| Treat | -0.580**** | -0.936**** |
|  | (0.026) | 0.031) |
| Treat × Total Daily Contacts |  | 0.203**** |
|  |  | (0.009) |
| Total Daily Contacts | 0.259**** | 0.153**** |
|  | (0.006) | (0.007) |
| Daily Travel Distance | 0.171**** | 0.166**** |
|  | (0.012) | (0.012) |
| Daily Avg. Travel Speed | 0.001**** | 0.001**** |
|  | (0.000) | (0.000) |
| Population Controlled | Yes | Yes |
| Block Land Area Controlled | Yes | Yes |
| Population Income Controlled | Yes | Yes |
| Population Gender Controlled | Yes | Yes |
| Population Race Controlled | Yes | Yes |
| Number of Existing Users Controlled | Yes | Yes |
| Number of Opt-in Users Controlled | Yes | Yes |
| Mobile App Usage Controlled | Yes | Yes |
| Week Fixed Effect | Yes | Yes |
| Day of Week Fixed Effect | Yes | Yes |
| Log likelihood | -33921.48 | -33692.75 |
| Observations | 26,500 | 26,500 |

\* $p < 0.10$, \*\* $p < 0.05$, \*\*\* $p < 0.01$, \*\*\*\* $p < 0.001$.

## Table B5 – Colorado Spring

| City = Colorado Spring | (1) | (2) |
|---|---|---|
| Tr. = National Emergency | Main Effect | Interaction w/ Daily Contacts |
| Treat | -0.678**** | 0.046 |
|  | (0.026) | (0.038) |
| Treat × Total Daily Contacts |  | -0.216**** |
|  |  | (0.008) |
| Total Daily Contacts | 0.192**** | 0.319**** |
|  | (0.005) | (0.007) |
| Daily Travel Distance | -0.559**** | -0.512**** |
|  | (0.015) | (0.015) |
| Daily Avg. Travel Speed | 0.135**** | 0.136**** |
|  | (0.002) | (0.002) |
| Population Controlled | Yes | Yes |
| Block Land Area Controlled | Yes | Yes |
| Population Income Controlled | Yes | Yes |
| Population Gender Controlled | Yes | Yes |
| Population Race Controlled | Yes | Yes |
| Number of Existing Users Controlled | Yes | Yes |
| Number of Opt-in Users Controlled | Yes | Yes |
| Mobile App Usage Controlled | Yes | Yes |
| Week Fixed Effect | Yes | Yes |
| Day of Week Fixed Effect | Yes | Yes |
| Log likelihood | 39011.67 | -38626.56 |
| Observations | 33,496 | 33,496 |

* $p < 0.10$, ** $p < 0.05$, *** $p < 0.01$, **** $p < 0.001$.

## Table B6 – Virginia Beach

| City = Virginia Beach<br>Tr. = National Emergency | (1)<br>Main Effect | (2)<br>Interaction w/ Daily Contacts |
|---|---|---|
| Treat | -0.543****<br>(0.029) | -0.966****<br>(0.039) |
| Treat ×<br>Total Daily Contacts |  | 0.136****<br>(0.008) |
| Total Daily Contacts | 0.386****<br>(0.005) | 0.311****<br>(0.007) |
| Daily Travel Distance | 0.038**<br>(0.018) | 0.043**<br>(0.018) |
| Daily Avg. Travel Speed | 0.105****<br>(0.003) | 0.104****<br>(0.003) |
| Population Controlled | Yes | Yes |
| Block Land Area Controlled | Yes | Yes |
| Population Income Controlled | Yes | Yes |
| Population Gender Controlled | Yes | Yes |
| Population Race Controlled | Yes | Yes |
| Number of Existing Users Controlled | Yes | Yes |
| Number of Opt-in Users Controlled | Yes | Yes |
| Mobile App Usage Controlled | Yes | Yes |
| Week Fixed Effect | Yes | Yes |
| Day of Week Fixed Effect | Yes | Yes |
| Log likelihood | -27880.47 | -27750.04 |
| Observations | 30,422 | 30,422 |

* $p < 0.10$, ** $p < 0.05$, *** $p < 0.01$, **** $p < 0.001$.

**Table B7 – SFO**

| City = SFO<br>Tr. = National Emergency | (1)<br>Main Effect | (2)<br>Interaction w/ Daily Contacts |
|---|---|---|
| Treat | -0.678**** | -0.673**** |
|  | (0.030) | (0.030) |
| Treat ×<br>Total Daily Contacts |  | -0.138* |
|  |  | (0.083) |
| Total Daily Contacts | 0.313**** | 0.322**** |
|  | (0.020) | (0.021) |
| Daily Travel Distance | 0.523**** | 0.525**** |
|  | (0.008) | (0.008) |
| Daily Avg. Travel Speed | 0.039**** | 0.039**** |
|  | (0.001) | (0.001) |
| Population Controlled | Yes | Yes |
| Block Land Area Controlled | Yes | Yes |
| Population Income Controlled | Yes | Yes |
| Population Gender Controlled | Yes | Yes |
| Population Race Controlled | Yes | Yes |
| Number of Existing Users Controlled | Yes | Yes |
| Number of Opt-in Users Controlled | Yes | Yes |
| Mobile App Usage Controlled | Yes | Yes |
| Week Fixed Effect | Yes | Yes |
| Day of Week Fixed Effect | Yes | Yes |
| Log likelihood | -53214.61 | -53213.16 |
| Observations | 132,924 | 132,924 |

* $p < 0.10$, ** $p < 0.05$, *** $p < 0.01$, **** $p < 0.001$.

**Table B8 – Jacksonville**

| City = Jacksonville<br>Tr. = National Emergency | (1)<br>Main Effect | (2)<br>Interaction w/ Daily Contacts |
|---|---|---|
| Treat | -0.254****<br>(0.030) | -0.612****<br>(0.035) |
| Treat ×<br>Total Daily Contacts |  | 0.225****<br>(0.011) |
| Total Daily Contacts | 0.319****<br>(0.007) | 0.226****<br>(0.008) |
| Daily Travel Distance | 0.254****<br>(0.011) | 0.257****<br>(0.011) |
| Daily Avg. Travel Speed | 0.089****<br>(0.001) | 0.088****<br>(0.001) |
| Population Controlled | Yes | Yes |
| Block Land Area Controlled | Yes | Yes |
| Population Income Controlled | Yes | Yes |
| Population Gender Controlled | Yes | Yes |
| Population Race Controlled | Yes | Yes |
| Number of Existing Users Controlled | Yes | Yes |
| Number of Opt-in Users Controlled | Yes | Yes |
| Mobile App Usage Controlled | Yes | Yes |
| Week Fixed Effect | Yes | Yes |
| Day of Week Fixed Effect | Yes | Yes |
| Log likelihood | -34270.20 | -34073.08 |
| Observations | 60,526 | 60,526 |

* $p < 0.10$, ** $p < 0.05$, *** $p < 0.01$, **** $p < 0.001$.

**Table B9 – New Orleans**

| City = New Orleans<br>Tr. = National Emergency | (1)<br>Main Effect | (2)<br>Interaction w/ Daily Contacts |
|---|---|---|
| Treat | -0.048*<br>(0.028) | -0.411****<br>(0.034) |
| Treat ×<br>Total Daily Contacts |  | 0.187****<br>(0.010) |
| Total Daily Contacts | 0.533****<br>(0.007) | 0.480****<br>(0.008) |
| Daily Travel Distance | 0.319****<br>(0.011) | 0.313****<br>(0.011) |
| Daily Avg. Travel Speed | 0.030****<br>(0.000) | 0.030****<br>(0.000) |
| Population<br>Controlled | Yes | Yes |
| Block Land Area<br>Controlled | Yes | Yes |
| Population Income<br>Controlled | Yes | Yes |
| Population Gender<br>Controlled | Yes | Yes |
| Population Race<br>Controlled | Yes | Yes |
| Number of Existing Users<br>Controlled | Yes | Yes |
| Number of Opt-in Users<br>Controlled | Yes | Yes |
| Mobile App Usage<br>Controlled | Yes | Yes |
| Week<br>Fixed Effect | Yes | Yes |
| Day of Week<br>Fixed Effect | Yes | Yes |
| Log likelihood | -31425.97 | -31258.84 |
| Observations | 45,156 | 45,156 |

* $p < 0.10$, ** $p < 0.05$, *** $p < 0.01$, **** $p < 0.001$.

**Table B10 – Omaha**

| City = Omaha<br>Tr. = National Emergency | (1)<br>Main Effect | (2)<br>Interaction w/ Daily Contacts |
|---|---|---|
| Treat | -0.392****<br>(0.028) | -0.628****<br>(0.035) |
| Treat ×<br>Total Daily Contacts |  | 0.101****<br>(0.009) |
| Total Daily Contacts | 0.335****<br>(0.005) | 0.283****<br>(0.007) |
| Daily Travel Distance | 0.144****<br>(0.013) | 0.145****<br>(0.013) |
| Daily Avg. Travel Speed | 0.109****<br>(0.002) | 0.109****<br>(0.002) |
| Population Controlled | Yes | Yes |
| Block Land Area Controlled | Yes | Yes |
| Population Income Controlled | Yes | Yes |
| Population Gender Controlled | Yes | Yes |
| Population Race Controlled | Yes | Yes |
| Number of Existing Users Controlled | Yes | Yes |
| Number of Opt-in Users Controlled | Yes | Yes |
| Mobile App Usage Controlled | Yes | Yes |
| Week Fixed Effect | Yes | Yes |
| Day of Week Fixed Effect | Yes | Yes |
| Log likelihood | -33362.12 | -33294.91 |
| Observations | 48,972 | 48,972 |

\* $p < 0.10$, ** $p < 0.05$, *** $p < 0.01$, **** $p < 0.001$.

**Table B11 – NYC**

| City = NYC<br>Tr. = National Emergency | (1)<br>Main Effect<br>(Possion) | (2)<br>Interaction w/ Daily Contacts<br>(Possion) |
|---|---|---|
| Treat | -0.344****<br>(0.027) | -0.493****<br>(0.028) |
| Treat ×<br>Total Daily Contacts |  | 0.305****<br>(0.015) |
| Total Daily Contacts | 0.373****<br>(0.008) | 0.312****<br>(0.008) |
| Daily Travel Distance | 0.541****<br>(0.008) | 0.521****<br>(0.008) |
| Daily Avg. Travel Speed | 0.053****<br>(0.001) | 0.053****<br>(0.001) |
| Time Trend | 0.023****<br>(0.001) | 0.024****<br>(0.001) |
| Control Variables |  |  |
| Population | Yes | Yes |
| Block Land Area | Yes | Yes |
| Population Income | Yes | Yes |
| Population Gender | Yes | Yes |
| Population Race | Yes | Yes |
| Number of Existing Users | Yes | Yes |
| Number of Opt-in Users | Yes | Yes |
| Mobile App Usage | Yes | Yes |
| Week Fixed Effect | Yes | Yes |
| Day of Week Fixed Effect | Yes | Yes |
| Log likelihood | -62927.55 | -62749.22 |
| Observations | 291,394 | 291,394 |

\* $p < 0.10$, \*\* $p < 0.05$, \*\*\* $p < 0.01$, \*\*\*\* $p < 0.001$.

**Table B12 – Pittsburgh**

| City = Pittsburgh<br>Tr. = National Emergency | (1)<br>Main Effect<br>(Possion) | (2)<br>Interaction w/ Daily Contacts<br>(Possion) |
|---|---|---|
| Treat | -0.457**** | -0.576**** |
|  | (0.029) | (0.031) |
| Treat ×<br>Total Daily Contacts |  | 0.151****<br>(0.014) |
| Total Daily Contacts | 0.300**** | 0.241**** |
|  | (0.007) | (0.009) |
| Daily Travel Distance | 0.329**** | 0.325**** |
|  | (0.009) | (0.009) |
| Daily Avg. Travel Speed | 0.063**** | 0.062**** |
|  | (0.001) | (0.001) |
| Time Trend | 0.041**** | 0.041**** |
|  | (0.001) | (0.001) |
| Control Variables |  |  |
| Population | Yes | Yes |
| Block Land Area | Yes | Yes |
| Population Income | Yes | Yes |
| Population Gender | Yes | Yes |
| Population Race | Yes | Yes |
| Number of Existing Users | Yes | Yes |
| Number of Opt-in Users | Yes | Yes |
| Mobile App Usage | Yes | Yes |
| Week Fixed Effect | Yes | Yes |
| Day of Week Fixed Effect | Yes | Yes |
| Log likelihood | -50332.35 | -50277.06 |
| Observations | 157,516 | 157,516 |

\* $p < 0.10$, \*\* $p < 0.05$, \*\*\* $p < 0.01$, \*\*\*\* $p < 0.001$.

**Table B13 – Oklahoma City**

| City = Oklahoma City<br>Tr. = National Emergency | (1)<br>Main Effect<br>(Possion) | (2)<br>Interaction w/ Daily Contacts<br>(Possion) |
|---|---|---|
| Treat | -0.188****<br>(0.027) | -0.365****<br>(0.030) |
| Treat ×<br>Total Daily Contacts |  | 0.153****<br>(0.011) |
| Total Daily Contacts | 0.262****<br>(0.006) | 0.193****<br>(0.008) |
| Daily Travel Distance | 0.250****<br>(0.010) | 0.246****<br>(0.010) |
| Daily Avg. Travel Speed | 0.087****<br>(0.001) | 0.087****<br>(0.001) |
| Time Trend | 0.027****<br>(0.001) | 0.027****<br>(0.001) |
| Control Variables |  |  |
| Population | Yes | Yes |
| Block Land Area | Yes | Yes |
| Population Income | Yes | Yes |
| Population Gender | Yes | Yes |
| Population Race | Yes | Yes |
| Number of Existing Users | Yes | Yes |
| Number of Opt-in Users | Yes | Yes |
| Mobile App Usage | Yes | Yes |
| Week Fixed Effect | Yes | Yes |
| Day of Week Fixed Effect | Yes | Yes |
| Log likelihood | -41041.46 | -40953.54 |
| Observations | 82,256 | 82,256 |

* $p < 0.10$, ** $p < 0.05$, *** $p < 0.01$, **** $p < 0.001$.

**Table B14 – Philadelphia**

| | (1)<br>Main Effect<br>(Possion) | (2)<br>Interaction w/ Daily Contacts<br>(Possion) |
|---|---|---|
| City = Philadelphia | | |
| Tr. = National Emergency | | |
| Treat | -0.476**** | -0.500**** |
| | (0.029) | (0.029) |
| Treat ×<br>Total Daily Contacts | | 0.141****<br>(0.021) |
| Total Daily Contacts | 0.240**** | 0.206**** |
| | (0.010) | (0.011) |
| Daily Travel Distance | 0.466**** | 0.462**** |
| | (0.007) | (0.007) |
| Daily Avg. Travel Speed | 0.043**** | 0.043**** |
| | (0.001) | (0.001) |
| Time Trend | 0.029**** | 0.029**** |
| | (0.001) | (0.001) |
| Control Variables | | |
| Population | Yes | Yes |
| Block Land Area | Yes | Yes |
| Population Income | Yes | Yes |
| Population Gender | Yes | Yes |
| Population Race | Yes | Yes |
| Number of Existing Users | Yes | Yes |
| Number of Opt-in Users | Yes | Yes |
| Mobile App Usage | Yes | Yes |
| Week Fixed Effect | Yes | Yes |
| Day of Week Fixed Effect | Yes | Yes |
| Log likelihood | -56737.51 | -56716.98 |
| Observations | 222,918 | 222,918 |

\* $p < 0.10$, \*\* $p < 0.05$, \*\*\* $p < 0.01$, \*\*\*\* $p < 0.001$.

**Table B15 – Austin**

| City = Austin<br>Tr. = National Emergency | (1)<br>Main Effect<br>(Possion) | (2)<br>Interaction w/ Daily Contacts<br>(Possion) |
|---|---|---|
| Treat | -0.312**** | -0.542**** |
|  | (0.027) | (0.031) |
| Treat ×<br>Total Daily Contacts |  | 0.172**** |
|  |  | (0.011) |
| Total Daily Contacts | 0.267**** | 0.202**** |
|  | (0.006) | (0.008) |
| Daily Travel Distance | 0.384**** | 0.376**** |
|  | (0.010) | (0.010) |
| Daily Avg. Travel Speed | 4.25e-07 | 5.15e-07 |
|  | (5.42e-07) | (5.42e-07) |
| Time Trend | 0.024**** | 0.024**** |
|  | (0.001) | (0.001) |
| Control Variables |  |  |
| Population | Yes | Yes |
| Block Land Area | Yes | Yes |
| Population Income | Yes | Yes |
| Population Gender | Yes | Yes |
| Population Race | Yes | Yes |
| Number of Existing Users | Yes | Yes |
| Number of Opt-in Users | Yes | Yes |
| Mobile App Usage | Yes | Yes |
| Week Fixed Effect | Yes | Yes |
| Day of Week Fixed Effect | Yes | Yes |
| Log likelihood | -41862.29 | -41756.93 |
| Observations | 65,190 | 65,190 |

* $p < 0.10$, ** $p < 0.05$, *** $p < 0.01$, **** $p < 0.001$.

## Table B16 – Seattle

| City = Seattle<br>Tr. = National Emergency | (1)<br>Main Effect<br>(Possion) | (2)<br>Interaction w/ Daily Contacts<br>(Possion) |
|---|---|---|
| Treat | -0.469****<br>(0.031) | -0.494****<br>(0.033) |
| Treat ×<br>Total Daily Contacts |  | 0.034****<br>(0.014) |
| Total Daily Contacts | 0.268****<br>(0.008) | 0.254****<br>(0.009) |
| Daily Travel Distance | 0.289****<br>(0.007) | 0.289****<br>(0.007) |
| Daily Avg. Travel Speed | 0.047****<br>(0.001) | 0.047****<br>(0.001) |
| Time Trend | 0.040****<br>(0.001) | 0.040****<br>(0.001) |
| Control Variables |  |  |
| Population | Yes | Yes |
| Block Land Area | Yes | Yes |
| Population Income | Yes | Yes |
| Population Gender | Yes | Yes |
| Population Race | Yes | Yes |
| Number of Existing Users | Yes | Yes |
| Number of Opt-in Users | Yes | Yes |
| Mobile App Usage | Yes | Yes |
| Week Fixed Effect | Yes | Yes |
| Day of Week Fixed Effect | Yes | Yes |
| Log likelihood | -60460.45 | -60457.51 |
| Observations | 297,542 | 297,542 |

\* $p < 0.10$, \*\* $p < 0.05$, \*\*\* $p < 0.01$, \*\*\*\* $p < 0.001$.

**Table B17 – Arlington**

| City = Arlington<br>Tr. = National Emergency | (1)<br>Main Effect<br>(Possion) | (2)<br>Interaction w/ Daily Contacts<br>(Possion) |
|---|---|---|
| Treat | -0.215****<br>(0.026) | -0.799****<br>(0.033) |
| Treat ×<br>Total Daily Contacts |  | 0.250****<br>(0.010) |
| Total Daily Contacts | 0.386****<br>(0.006) | 0.258****<br>(0.008) |
| Daily Travel Distance | 0.056****<br>(0.016) | 0.054***<br>(0.016) |
| Daily Avg. Travel Speed | 0.071****<br>(0.002) | 0.071****<br>(0.002) |
| Time Trend | 0.033****<br>(0.001) | 0.034****<br>(0.001) |
| Control Variables |  |  |
| Population | Yes | Yes |
| Block Land Area | Yes | Yes |
| Population Income | Yes | Yes |
| Population Gender | Yes | Yes |
| Population Race | Yes | Yes |
| Number of Existing Users | Yes | Yes |
| Number of Opt-in Users | Yes | Yes |
| Mobile App Usage | Yes | Yes |
| Week Fixed Effect | Yes | Yes |
| Day of Week Fixed Effect | Yes | Yes |
| Log likelihood | -26131.21 | -25918.39 |
| Observations | 27,772 | 27,772 |

* $p < 0.10$, ** $p < 0.05$, *** $p < 0.01$, **** $p < 0.001$.

**Table B18 – Phoenix**

| City = Phoenix<br>Tr. = National Emergency | (1)<br>Main Effect<br>(Possion) | (2)<br>Interaction w/ Daily Contacts<br>(Possion) |
|---|---|---|
| Treat | -0.441****<br>(0.030) | -0.493****<br>(0.033) |
| Treat ×<br>Total Daily Contacts |  | 0.077****<br>(0.015) |
| Total Daily Contacts | 0.245****<br>(0.008) | 0.218****<br>(0.010) |
| Daily Travel Distance | 0.384****<br>(0.007) | 0.383****<br>(0.007) |
| Daily Avg. Travel Speed | 5.85e-07***<br>(1.90e-07) | 5.91e-07****<br>(1.90e-07) |
| Time Trend | 0.036****<br>(0.001) | 0.037****<br>(0.001) |
| Control Variables |  |  |
| Population | Yes | Yes |
| Block Land Area | Yes | Yes |
| Population Income | Yes | Yes |
| Population Gender | Yes | Yes |
| Population Race | Yes | Yes |
| Number of Existing Users | Yes | Yes |
| Number of Opt-in Users | Yes | Yes |
| Mobile App Usage | Yes | Yes |
| Week Fixed Effect | Yes | Yes |
| Day of Week Fixed Effect | Yes | Yes |
| Log likelihood | -60476.09 | -60464.22 |
| Observations | 264,364 | 264,364 |

\* $p < 0.10$, \*\* $p < 0.05$, \*\*\* $p < 0.01$, \*\*\*\* $p < 0.001$.

**Table B19 – Nashville**

| City = Nashville<br>Tr. = National Emergency | (1)<br>Main Effect<br>(Possion) | (2)<br>Interaction w/ Daily Contacts<br>(Possion) |
|---|---|---|
| Treat | -0.264**** | -0.406**** |
|  | (0.027) | (0.029) |
| Treat ×<br>Total Daily Contacts |  | 0.203**** |
|  |  | (0.014) |
| Total Daily Contacts | 0.279**** | 0.222**** |
|  | (0.007) | (0.009) |
| Daily Travel Distance | 0.363**** | 0.354**** |
|  | (0.009) | (0.010) |
| Daily Avg. Travel Speed | 0.048**** | 0.048**** |
|  | (0.001) | (0.001) |
| Time Trend | 0.025**** | 0.025**** |
|  | (0.001) | (0.001) |
| Control Variables |  |  |
| Population | Yes | Yes |
| Block Land Area | Yes | Yes |
| Population Income | Yes | Yes |
| Population Gender | Yes | Yes |
| Population Race | Yes | Yes |
| Number of Existing Users | Yes | Yes |
| Number of Opt-in Users | Yes | Yes |
| Mobile App Usage | Yes | Yes |
| Week Fixed Effect | Yes | Yes |
| Day of Week Fixed Effect | Yes | Yes |
| Log likelihood | -36856.10 | -36760.59 |
| Observations | 58,512 | 58,512 |

\* $p < 0.10$, \*\* $p < 0.05$, \*\*\* $p < 0.01$, \*\*\*\* $p < 0.001$.

**Table B20 – Wichita**

| City = Wichita<br>Tr. = National Emergency | (1)<br>Main Effect<br>(Possion) | (2)<br>Interaction w/ Daily Contacts<br>(Possion) |
|---|---|---|
| Treat | -0.239**** | -0.474**** |
|  | (0.027) | (0.039) |
| Treat ×<br>Total Daily Contacts |  | 0.076**** |
|  |  | (0.009) |
| Total Daily Contacts | 0.252**** | 0.214**** |
|  | (0.006) | (0.007) |
| Daily Travel Distance | -0.015 | -0.015 |
|  | (0.016) | (0.017) |
| Daily Avg. Travel Speed | 0.129**** | 0.127**** |
|  | (0.002) | (0.002) |
| Time Trend | 0.031**** | 0.031**** |
|  | (0.001) | (0.001) |
| Control Variables |  |  |
| Population | Yes | Yes |
| Block Land Area | Yes | Yes |
| Population Income | Yes | Yes |
| Population Gender | Yes | Yes |
| Population Race | Yes | Yes |
| Number of Existing Users | Yes | Yes |
| Number of Opt-in Users | Yes | Yes |
| Mobile App Usage | Yes | Yes |
| Week Fixed Effect | Yes | Yes |
| Day of Week Fixed Effect | Yes | Yes |
| Log likelihood | -29658.793 | -29626.42 |
| Observations | 33,072 | 33,072 |

* $p < 0.10$, ** $p < 0.05$, *** $p < 0.01$, **** $p < 0.001$.

# Appendix C

## Table C – Robustness Test with Controls for Spatially Adjacent Blocks

| City = DC<br>Tr. = National Emergency | (1)<br>Main Effect<br>(Possion) | (2)<br>Interaction w/ Daily Contacts<br>(Possion) |
|---|---|---|
| Treat | -0.199**** | -0.322**** |
|  | (0.028) | (0.030) |
| Treat ×<br>Total Daily Contacts |  | 0.145**** |
|  |  | (0.012) |
| Total Daily Contacts | 0.214**** | 0.173**** |
|  | (0.007) | (0.007) |
| Daily Travel Distance | 0.298**** | 0.289**** |
|  | (0.011) | (0.011) |
| Daily Avg. Travel Speed | 0.071**** | 0.072**** |
|  | (0.001) | (0.001) |
| Time Trend | 0.015**** | 0.015**** |
|  | (0.001) | (0.001) |
| *Control Variables* |  |  |
| Focal Block Population | Yes | Yes |
| Focal Block Land Area | Yes | Yes |
| Focal Block Population Income | Yes | Yes |
| Focal Block Population Gender | Yes | Yes |
| Focal Block Population Race | Yes | Yes |
| Focal Block Number of Existing Users | Yes | Yes |
| Focal Block Number of Opt-in Users | Yes | Yes |
| Focal Block Mobile App Usage | Yes | Yes |
| 1st Closest Block Population | Yes | Yes |
| 1st Closest Block Land Area | Yes | Yes |
| 1st Closest Block Population Income | Yes | Yes |
| 1st Closest Block Population Gender | Yes | Yes |
| 1st Closest Block Population Race | Yes | Yes |
| 1st Closest Block Number of Existing Users | Yes | Yes |
| 1st Closest Block Number of Opt-in Users | Yes | Yes |
| 1st Closest Block Mobile App Usage | Yes | Yes |
| 2nd Closest Block Population | Yes | Yes |
| 2nd Closest Block Land Area | Yes | Yes |
| 2nd Closest Block Population Income | Yes | Yes |
| 2nd Closest Block Population Gender | Yes | Yes |
| 2nd Closest Block Population Race | Yes | Yes |
| 2nd Closest Block Number of Existing Users | Yes | Yes |
| 2nd Closest Block Number of Opt-in Users | Yes | Yes |
| 2nd Closest Block Mobile App Usage | Yes | Yes |
| 3rd Closest Block Population | Yes | Yes |
| 3rd Closest Block Land Area | Yes | Yes |
| 3rd Closest Block Population Income | Yes | Yes |
| 3rd Closest Block Population Gender | Yes | Yes |
| 3rd Closest Block Population Race | Yes | Yes |
| 3rd Closest Block Number of Existing Users | Yes | Yes |
| 3rd Closest Block Number of Opt-in Users | Yes | Yes |
| 3rd Closest Block Mobile App Usage | Yes | Yes |
| Week Fixed Effect | Yes | Yes |
| Day of Week Fixed Effect | Yes | Yes |
| Log likelihood | -41420.74 | -41352.1 |
| Observations | 84,270 | 84,270 |

\* $p < 0.10$, \*\* $p < 0.05$, \*\*\* $p < 0.01$, \*\*\*\* $p < 0.001$.

**Appendix D**

**Table D – COVID-19 Health Risks in Blue Cities versus Red Cities**

|  | DV = Infection Rate | DV = Death Rate |
|---|---|---|
| Blue City | 0.046**** | 0.011**** |
|  | (0.000) | (0.000) |
| R-squared | 0.031 | 0.022 |
| Observations | 1,767,338 | 1,767,338 |

* $p < 0.10$, ** $p < 0.05$, *** $p < 0.01$, **** $p < 0.001$.

**Appendix E – Additional Falsification Test**

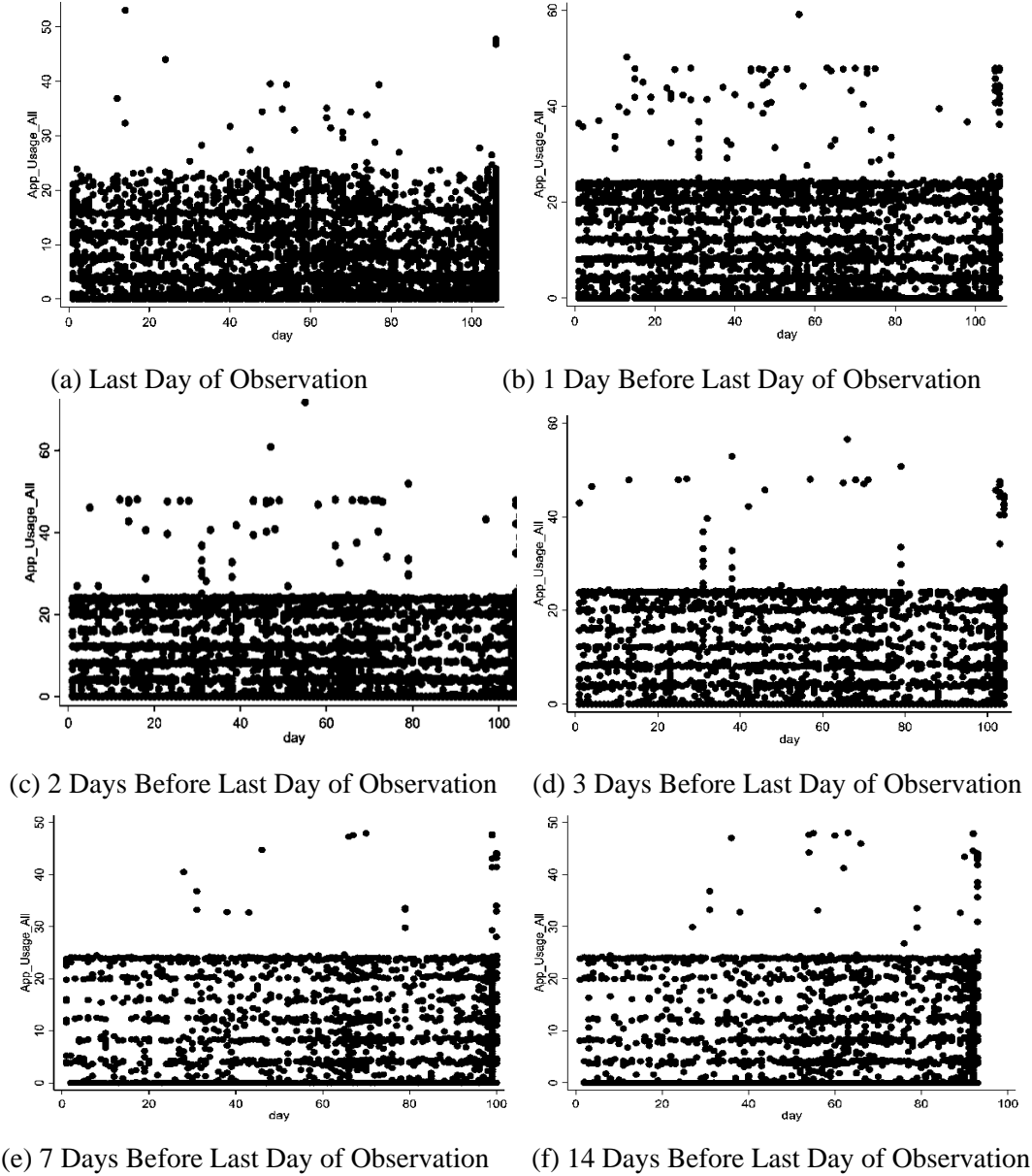

(a) Last Day of Observation  (b) 1 Day Before Last Day of Observation

(c) 2 Days Before Last Day of Observation  (d) 3 Days Before Last Day of Observation

(e) 7 Days Before Last Day of Observation  (f) 14 Days Before Last Day of Observation

**Figure E. Distribution of total #app usage time (minutes) for each "opt-out" user on (a) the last day of observing that user; (b) 1 day before last day; (c) 2 days before last day; (d) 3 days before last day; and (e) 1 week before last day.**